\documentclass[12pt]{article} 
\pdfoutput=1 
\usepackage{jcapmod} 
\usepackage{booktabs}
\usepackage{mathtools}
\usepackage[english]{babel}
\usepackage{amsmath,amssymb,amsbsy,amstext, amsthm}
\usepackage[usenames,dvipsnames]{xcolor}
\usepackage{hyperref}
\hypersetup{
    urlcolor=NavyBlue,
    citecolor=NavyBlue,
    linkcolor=NavyBlue,
}%
\usepackage{array}
\usepackage{url} 
\usepackage{slashed}
\usepackage{multicol}
\usepackage{blkarray}
\usepackage{enumerate}
\usepackage{graphicx}
\usepackage{amsfonts}
\usepackage{amssymb}
  \usepackage{empheq}
  \usepackage{tcolorbox}
\usepackage{xcolor}
\usepackage{soul}
\usepackage{colortbl}

\usepackage{float}
\usepackage[utf8]{inputenc}
\RequirePackage{color}
\usepackage[normalem]{ulem}
\def\be{\begin{equation}}
\def\ee{\end{equation}}
\definecolor{verde}{rgb}{0,0.5,0}

\def\bea{\begin{eqnarray}}
\def\eea{\end{eqnarray}}
\def\nn{\nonumber}

\title{On the prospects of thermalization of axion-SU(2) inflation}

\author[\spadesuit,1]{Sukannya Bhattacharya, \note{corresponding author}}

\author[\spadesuit]{Matteo Fasiello,}
\author[\spadesuit]{Alexandros Papageorgiou,}
\author[\diamondsuit]{and Ema Dimastrogiovanni}

\affiliation[\diamondsuit]{Van Swinderen Institute for Particle Physics and Gravity, University of Groningen, Nijenborgh 4, 9747 AG Groningen, The Netherlands}
\affiliation[\spadesuit]{Instituto de F\'{i}sica T\'{e}orica UAM-CSIC, c/ Nicol\'{a}s Cabrera 13-15, 28049, Madrid, Spain}

\emailAdd{sukannya.bhattacharya@ift.csic.es}
\emailAdd{matteo.fasiello@csic.es}
\emailAdd{papageorgiou.hep@gmail.com}
\emailAdd{e.dimastrogiovanni@rug.nl}

\date{March 2025}

\abstract{Axion inflation models coupled to a gauge sector via a Chern-Simons term exhibit an array of interesting phenomenology including a chiral gravitational wave spectrum and primordial black hole production. They may also provide a useful mechanism for generating lepton asymmetry. The possibility to embed this class of models in UV-finite theories and their intriguing, testable, signatures make for a compelling candidate for early acceleration.
 Due to the Chern-Simons coupling, gauge modes may undergo a finite tachyonic growth during which non-linearities become important. Naturally, this raises the question of whether such (self) interactions can lead to thermalization during inflation.
We provide a set of useful criteria for sustained thermalization in an  axion-$SU(2)$ model and chart the parameter space of the model accordingly. We find that the cold inflation regime constitutes  a very significant fraction of the parameter space. Our analysis accounts for a initially vanishing as well as non-zero gauge field vacuum expectation value (VEV). We also consider the possibility of a dynamically generated  VEV.}

\begin{document}

\maketitle
\section{Introduction}

Ever since its proposal, natural inflation \cite{Freese:1990rb,Adams:1992bn}—sometimes referred to simply as axion inflation—has held a special status as one of the leading candidates for a successful realization of the primordial inflationary scenario. The reasons are manifold. First, the axion-inflaton enjoys an approximate shift symmetry that protects its potential from large radiative corrections; i.e. it solves the $\eta$-problem \cite{Copeland:1994vg}. 
Secondly, string theory generically predicts the presence of multiple axions at ``lower'' energies as a direct result of string compactifications \cite{Baumann:2014nda}. Such a scenario opens the door for some of these axions to have just the right properties for a successful inflation. Finally, axions are generally expected to couple to gauge fields through a dimension-5 operator of the form $\phi F \tilde{F}$ where $\phi$ is the axion field and $F$ is the gauge field strength tensor. The Chern-Simons type coupling leads to a rich phenomenology, especially at small scales, thus providing a unique  opportunity to test these models in the near future. We refer the interested reader to \cite{Pajer:2013fsa} for a post-Planck review of axion inflation models.

On the observational side, one ought to first take into account the fact that the simplest realization of axion inflation is at odds with the latest CMB measurements \cite{Planck:2018jri,BICEP:2021xfz}. The slope of the axion potential is inversely proportional to the axion decay constant $f$, so that large $f$ values are needed to reproduce the nearly flat spectrum of scalar fluctuations. On the other hand, the axion decay constant is typically found to be sub-Planckian in UV-complete theories such as string theory \cite{Banks:2003sx}. There is a host of interesting and well-motivated mechanisms to effectively flatten the axion potential \cite{Arkani-Hamed:2003xts,Kim:2004rp,Dimopoulos:2005ac,Silverstein:2008sg,McAllister:2008hb,Kaloper:2008fb,Marchesano:2014mla,Bachlechner:2014gfa,Choi:2014rja,Kappl:2015esy,Parameswaran:2016qqq,Abe:2014xja} but it remains challenging for  minimal models to account for CMB observations. 

An interesting possibility in this context is to consider an additional\footnote{Additional here with respect to the ever-present Hubble friction.} source of friction leading to a slow-down of the axion even for a steep potential. The presence of such a term typically allows for a smaller axion decay constant while ensuring compatibility with CMB constraints. 
Such scenarios generally fall into two categories:~(i) non-minimal coupling of the axion with gravity \cite{Watanabe:2020ctz,Almeida:2020kaq,Dimastrogiovanni:2023oid}, with the  slow down of the axion as a direct result of the gravitational background; (ii) friction is obtained  by means of an interaction between the axion and a gauge sector during inflation. The latter possibility has been extensively studied in the case of both Abelian \cite{Anber:2009ua,Barnaby:2011vw,Barnaby:2010vf,Cook:2011hg,Ozsoy:2014sba,Cheng:2015oqa,Peloso:2016gqs,Domcke:2020zez,Gorbar:2021rlt,Peloso:2022ovc,Garcia-Bellido:2023ser,vonEckardstein:2023gwk,Durrer:2024ibi,Corba:2024tfz,Alam:2024fid,He:2025ieo,Jimenez:2017cdr} and non-Abelian \cite{Maleknejad:2011sq,Adshead:2012kp,Dimastrogiovanni:2012ew,Dimastrogiovanni:2012st,Adshead:2013nka,Obata:2016tmo,Adshead:2016omu,Holland:2020jdh,Ishiwata:2021yne,Iarygina:2023mtj,Putti:2024uyr} gauge fields. Intriguingly, this configuration has been shown to support a significant  production of helical gauge fields which may in turn  source gravitational waves \cite{Sorbo:2011rz,Barnaby:2012xt,Mukohyama:2014gba,Namba:2015gja,Garcia-Bellido:2016dkw,Dimastrogiovanni:2016fuu,Ozsoy:2017blg,Thorne:2017jft,Unal:2023srk,Ozsoy:2024apn,Bastero-Gil:2022fme}, support large non-Gaussianity \cite{Agrawal:2017awz,Ozsoy:2021onx} and primordial black holes production \cite{Garcia-Bellido:2017aan,Dimastrogiovanni:2024xvc}, explain the lepton number asymmetry \cite{Caldwell:2017chz,Papageorgiou:2017yup,Domcke:2019mnd} and the origin of primordial magnetic fields \cite{Adshead:2016iae,Sobol:2019xls,Brandenburg:2024awd}. 
Given that the gauge field dynamics can become significantly non-linear, it is important to probe the evolution of background and fluctuations via  non-perturbative methods such as lattice simulations. Several lattice studies of the coupling of axion and U(1) gauge field already populate the literature \cite{Caravano:2021bfn,Caravano:2022epk,Figueroa:2023oxc,Figueroa:2024rkr,Caravano:2024xsb,Sharma:2024nfu}. 

From this point forward, we shall focus on models of axion inflation in the presence of non-Abelian gauge fields. The action in question takes the simple form
\begin{equation}
    S=\int {\rm d}^4 x\sqrt{-g}\left[\frac{M_{\rm Pl}^2}{2} R-\frac{1}{2}\left(\partial\phi\right)^2-V(\phi)-\frac{1}{4}F^a_{\;\mu\nu} F^{a\,\mu\nu}+\frac{\lambda \phi}{4 f} F^a_{\;\mu\nu}\tilde{F}^{a\,\mu\nu}\right]\;,
    \label{CNIL}
\end{equation}
with minimal coupling to gravity and a Chern-Simons interaction between an axion-inflaton and a non-Abelian gauge sector. 
In the last few years, the same underlying model has been the subject of several studies focusing broadly on two rather different regimes.
One set of works takes on the above Chromo-Natural Inflation (CNI) Lagrangian  postulating the existence (or showing the emergence \cite{Domcke:2018rvv,Domcke:2019lxq}) of a non-zero, isotropic gauge field background with field perturbations that feature a cold (or, in other words, completely  non-thermal) spectrum. Another, equally interesting and somewhat complementary, set of works \cite{Berghaus:2019whh,DeRocco:2021rzv,Berghaus:2024zfg,Kamali:2024qme,Kamali:2019ppi,Fujita:2025zoa} starts out with a zero gauge field background and a thermal spectrum for  gauge field perturbations. This is a particular realization of the warm inflation paradigm which is well studied in the literature~\cite{Berera:1995ie,Ramos:2013nsa,Berera:2008ar,Bastero-Gil:2009sdq,Bartrum:2013fia,Bastero-Gil:2016qru,Taylor:2000ze,Berera:1998gx,Bastero-Gil:2011rva,Kamali:2023lzq}. Only very recently there have been efforts to study the two regimes in tandem \cite{Mukuno:2024yoa}.

The purpose of the present work is to shed some light on the conditions under which thermalization may occur  and approximately delineate the parameter space which naturally belongs to the cold  \textit{vs} the warm inflation regime. To this end, we study an initial setup where  perturbations are in their Bunch-Davies vacuum configuration   and examine two CNI scenarios, one with a zero initial gauge field background and one with a non-vanishing background.  To determine the importance of non-linearities, one should study their impact on the otherwise  standard (linear) set of equations of motion in CNI. The threshold corresponding to non-linear effects that may no longer be neglected can be identified by providing evidence of (i) efficient scattering in a particle-like description and/or (ii) violation of perturbativity bounds for the gauge field spectrum. In this work we shall pursue both of these characterizations. 

First, we will adopt a particle-like description for perturbations and consider the Boltzmann equation for the gauge field occupation number. Our analysis follows the approach adopted in \cite{Ferreira:2017lnd}, which features an  Abelian gauge sector. A particle-like description is suitable when the wavelength of the perturbations is sufficiently smaller than the size of the horizon. In our setup, it turns out to be a rather convenient approach as it combines the effects of both cubic and quartic self-couplings of the gauge field in a consistent manner. This procedure entails computing the scattering rate of gauge field perturbations and identifying where it becomes the dominant effect in the Boltzmann equation. The second approach relies on assuming that perturbativity bounds are a good proxy for (the onset of) thermalization. We track the breakdown of perturbativity by computing the ratio of the 1-loop power spectrum of gauge field perturbations with the tree-level result. The validity of this last approach will be justified a posteriori.

The presence of large non-linearities does not itself guarantee thermalization but merely indicates that there is efficient exchange of energy between the gauge  modes. 
For a large scattering rate one would expect the spectrum to transition from a ``cold" to a ``warm" regime as a result of the second law of thermodynamics \cite{DeRocco:2021rzv}. However, showing clear evidence of the transition into a fully thermal configuration (e.g. characterized by a temperature parameter $T$),  especially in a rapidly expanding universe, requires sophisticated lattice simulations which are beyond the scope of this work. Instead, we will provide here a  chart of the parameter space for which thermalization is within reach\footnote{Mind that this way of organizing the parameter space can be seen as a ``conservative'' approach by those interested in working safely within the cold inflation regime. However, we stress here that working outside these parameter space limits does \textit{not} guarantee thermalization.}.
Once established that interactions are highly non-linear, to reach  thermal equilibrium one must ensure that the cold spectrum transitions into a thermal spectrum and that this transition is robust against the quasi-exponential expansion taking place during inflation. 
 
This paper is organized as follows. In section \ref{sec:background} we give a brief overview of the  axion-$SU(2)$ inflationary model with and without an isotropic gauge field background. In section \ref{sec:interactions} we calculate the efficiency of the interaction rate among the modes. We first adopt  the particle description and then compare our findings on the onset of warm inflation with those stemming from perturbativity limits. In section \ref{sec:conditions} we formalize the conditions for thermalization. In section \ref{sec:results} we discuss our key results and place them in the broader context. In section~\ref{sec:conclusions} we compare our findings with the existing literature on the subject and draw our conclusions.
We supplement  the paper with several Appendices containing details  that are best avoided in the main text.

\section{Brief review of axion-$SU(2)$ inflation}
\label{sec:background}

\subsection{Background} 

We start our analysis with a brief overview of axion-$SU(2)$ inflation model at the background level. The Lagrangian is given in eq.~(\ref{CNIL}). 
One may readily show \cite{Maleknejad:2011jw} the existence of an isotropic background\footnote{Studies exist \cite{Maleknejad:2013npa} supporting the claim that such isotropic solution is indeed an attractor.} obeying the following ansatz
 
\be
 A^a_0 (t) = 0, ~ ~ ~ ~  A^a_i (t) = \delta ^a _i a(t) Q(t),\label{eq:backgauge}
\ee
where the lower indices $0$ and $i=1,2,3$ are for time and space, and upper indices $a=1,2,3$ are for the gauge group. We reserve the symbol $a(t)$ for the scale factor and assume a flat FLRW metric with mostly positive diagonal components. The self-coupling of the non-Abelian gauge field is denoted by $g$, the axion decay constant is $f$ and  $\lambda$ is a dimensionless parameter that quantifies the strength of the Chern-Simons coupling.

Variation of the action with respect to the 00 and ii components of the metric yield the first and second Friedmann equation, respectively
\begin{align}
    3 H^{2} M_{\rm Pl}^{2}=&
    \frac{\dot{\phi}^{2}}{2}+V(\phi)+\frac{3}{2}\Big[\left(\dot{Q}+HQ \right)^{2}+g^{2}Q^{4}\Big]\,,\label{eq:H}\\
    -2M_{\rm Pl}^{2}\dot{H}=&
    \dot{\phi}^{2}+2\left[\left(\dot{Q}+HQ \right)^{2}+g^{2}Q^{4}\right]\,.\label{eq:Hdot}
\end{align}
Additionally, variation of the action with respect to the background fields $\phi(t)$ and $Q(t)$ results in the field equations of motion
\begin{align}
    \ddot{\phi}+3H\dot{\phi}&+V_{\phi}+\frac{3 g \lambda}{f}Q^{2}\left(\dot{Q}+H Q\right)=0\,,\label{eq:eqchi}\\
    \ddot{Q}+3H\dot{Q}&+\left(\dot{H}+2H^{2} \right)Q+gQ^{2}\left(2gQ-\frac{\lambda \dot{\phi}}{f} \right)=0\,.\label{eq:eqQ}
\end{align}
Note that the above expressions neglect  backreaction contributions \cite{Iarygina:2023mtj,Dimastrogiovanni:2024xvc,Brandenburg:2024awd,Dimastrogiovanni:2025snj} and hence the results derived in this work will be valid in the weak-backreaction regime of axion-$SU(2)$ inflation. 
The parameter space regime corresponding to weak backreaction is well known for this class of models \cite{Dimastrogiovanni:2016fuu,Maleknejad:2018nxz} so that so we can retroactively check the soundness of our  assumption.

The above system of equations is the most general background setup for axion-$SU(2)$ inflation in the low backreaction regime. One recovers the case of conventional slow-roll inflation upon setting $Q(t)=0$. On the other hand, if we assume the background gauge field to be non-zero, there is a well-known solution that allows for the axion-inflaton to roll down a steep potential because of the additional friction induced by the gauge field. In turn, the gauge field remains nearly constant as it is sourced by the axion. This regime can be arrived at from the equations of motion by imposing the slow-roll conditions $\ddot{\phi}\simeq\ddot{Q}\simeq0$ while simultaneously assuming the following hierarchies in the parameters, $\lambda Q/f\gg2$ and $\lambda\, g\, Q^2/(f H)\gg \sqrt{3}$. The system simplifies to
\begin{align}
    &V_{\phi}+\frac{3 g \lambda H}{f}Q^{3}\simeq 0\,,\\
    &2H^{2}Q+gQ^{2}\left(2gQ-\frac{\lambda \dot{\phi}}{f} \right)\simeq 0\,,
\end{align}
These expressions can be recast into the ``master" formulas that describe the weak backreaction regime in chromo-natural inflation:
\begin{equation}
    Q\simeq\left(\frac{-f V_\phi}{3 g \lambda H}\right)^{1/3}\;\;\;,\;\;\;\xi\simeq m_Q+\frac{1}{m_Q}\;,
    \label{eq:CNI-attractor}
\end{equation}
where we have defined the two parameters
\begin{equation}
    m_Q\equiv \frac{g Q}{H} \;\;\;,\;\;\;\xi\equiv \frac{\lambda \dot{\phi}}{2H f} \;.
\end{equation}
These two dimensionless quantities are sometimes referred to as ``particle production parameters" in the literature. They track the gauge field background value and the speed of the axion respectively. The same two parameters control the rate of production of perturbations during inflation. Of course, in the regime of vanishing gauge field background, one has $m_Q=0$ while $\xi$ remains non-zero. When the gauge field background is non-zero, the two parameters enjoy a one-to-one relation that is well approximated by eq.~(\ref{eq:CNI-attractor}).

\subsection{Perturbations}

The  perturbations of the non-Abelian gauge field, $T_{ai} \subset \delta A^a_i$ differ significantly contingent on whether the gauge field background is zero or not. Focusing now on the $Q(t)=0$ case, the most natural decomposition of the perturbation consists of three copies of Abelian-like gauge fields. The perturbations are transverse and hence the six degrees of freedom can be expressed as the two circular polarizations of the three copies of the Abelian-like vector fields. In Fourier space we write

\begin{equation}
    T^a_{{\cal AB,\;}i}(\tau,\vec{x})=\int \frac{d^3 q}{(2\pi)^{3/2}}{e}^{i\vec{q}\vec{x}}\sum_{\sigma=\pm}\, \epsilon_{i,\sigma}(\vec{q})T^a_{\cal AB,\;\sigma}(\vec{q},\tau)
\end{equation}

where the subscript ${\cal AB}$ signifies the Abelian regime perturbations of the non-Abelian theory and the polarization vectors $\epsilon _{\pm}(\vec{q})$ satisfy the following relations
\bea
\vec{q}.\vec{\epsilon}_{\pm}(\vec{q})&=&0 ~ ~ ~ ~ \vec{q}\times \vec{\epsilon}_{\pm}(\vec{q})=\mp iq \vec{\epsilon}_{\pm}(\vec{q}), \nonumber \\
\vec{\epsilon}_{ \sigma}(\vec{q})\vec{\epsilon}_{\sigma '}(\vec{q})^{\star} &=& \delta _{\sigma \sigma '} ~ ~ ~ ~ \vec{\epsilon}_{\pm}(\vec{q})^{\star} = \vec{\epsilon}_{\pm}(-\vec{q})= \vec{\epsilon}_{\mp}(\vec{q})
\eea
The equations of motion then reduce to the well known expressions for Abelian gauge field perturbations \cite{Anber:2009ua,Barnaby:2010vf}
\begin{equation}
    \frac{d^2T_{{\cal AB},\,{\pm}}^a(\vec{k},\tau)}{dx^2} + \bigg[1\mp \frac{2\xi}{x}\bigg]T_{{\cal AB,\,}{\pm}}^a(\vec{k},\tau)=0, ~ ~ ~ x\equiv -k\tau .
\end{equation}
The solution for the tachyonically enhanced polarization is well approximated in terms of a Whittaker function
\begin{equation}
    T_{{\cal AB},\,+}^a(\tau,k)\simeq \frac{e^{\frac{\pi}{2}\xi}}{\sqrt{2k}}W_{-i\xi, 1/2}(-2i x),\label{eq:Abelianlimit_Wh}
\end{equation}
true for $a=1,2,3$.
For a constant $\xi$, the above solution provides a good approximation to account for the enhancement of a particular mode inside the horizon within the band $(8\xi)^{-1}<x<2\xi$, see~\cite{Barnaby:2010vf,Barnaby:2011vw}.

In the presence of a non-zero gauge field background, 
it is easy to see that out of the original six degrees of freedom in the gauge sector, two couple linearly to standard tensor modes. We shall focus only on these two as they feature a well known tachyonic instability. In Fourier space one has 
\be
T_{ab}(\tau, \vec{x})= \int \frac{d^3q}{(2\pi)^{3/2}}e^{i\vec{q}\vec{x}}\sum _{\sigma =\pm} \epsilon _{a,\sigma}(\vec{q})\epsilon _{b,\sigma}(\vec{q})T_{\sigma}(\vec{q},\tau),\label{eq:Tft}\; .
\ee

From the action in Eq.~\eqref{CNIL}, the equation of motion for the canonically normalized transverse and traceless component of the gauge field is
\be
\frac{d^2T_{\pm}(\vec{k},\tau)}{dx^2} + \bigg[1+ \frac{2m_Q\xi}{x^2}\mp \frac{2(\xi +m_Q)}{x}\bigg]T_{\pm}(\vec{k},\tau)=0,
\label{eq:Tfree}
\ee
where ``$\pm$'' indicates the two polarizations . Here we have neglected possible linear mixing with the metric perturbations, i.e. source terms of $\mathcal{O}(h_{\pm})$. Such contributions are suppressed by powers of $M_{\rm Pl}$ and they do not affect the perturbations except very far outside the horizon \cite{Adshead:2013nka}. 

The $T_+$ polarization acquires a transient tachyonic mass and grows exponentially in the range $x_-< x < x_+$, where 
\be
x_{\pm} = m_Q+\xi \pm \sqrt{m_Q^2+\xi ^2}.
\label{eq:xpm}
\ee
In terms of $x$, the moment when the mode grows the fastest, is the moment when the tachyonic mass attains its greatest  negative value. We define this  as the ``moment of maximum instability", $x_m$. Additionally, $x_-$ corresponds to the moment when the instability switches off (always inside the horizon). As a result, we expect the amplitude of the perturbations to stop growing at this moment. We will call it the ``moment of maximum amplitude".
 
Assuming slow-roll evolution such that $\xi$ and $m_Q$ are nearly constant during the tachyonic evolution of a given mode, the above Eq.~\eqref{eq:Tfree} has an analytical solution for $T_+$ in terms of a Whittaker function:
\be
T_+(\vec{k},\tau) = \frac{e^{\frac{\pi}{2}(m_Q+\xi)}}{\sqrt{2k}}W_{\alpha, \beta}(2ik\tau), ~ ~ {\rm where} ~ ~\alpha = -i(\xi + m_Q),  ~ \beta = -i\sqrt{2m_Q^2 +\frac{7}{4}}\;.\label{Tfreesol}
\ee
In the next sections, we will often use this  solution since $\xi$ and $m_Q$ are indeed constant for the duration of evolution of our interest. In the case of zero vev Eq.~\eqref{Tfreesol} approaches the solution for the case of Abelian gauge fields, with $\alpha = -i\xi$ and $\beta = 1/2$. In such configuration the  tachyonic growth begins at $x=2\xi$, and persists beyond horizon crossing ($x=1$), the point of maximum instability being at $x_{m,0}=\xi$.
 
\subsection{Nonlinear interactions}

Expanding the last two terms in Eq.~\eqref{CNIL} we obtain cubic and quartic gauge field self-interactions
\bea
\mathcal{L}^{(3)}& \supset & -gf^{abc}T_{ai}T_{bj}\partial _iT_{cj} - \frac{g\xi }{3\tau} f^{abc}\epsilon ^{ijk}T_{ai}T_{bj}T_{ck} -\frac{gm_Q}{\tau}T_{ij}T_{jk}T_{ki}  \\
\mathcal{L}^{(4)} &\supset & -\frac{1}{4}g^2f^{abc}f^{ade}T_{bi}T_{cj}T_{di}T_{ej} .
\label{eq:L3L4}
\eea 
Interactions where two gauge modes in the initial state lead to two gauge modes in the final state will receive contributions from $\mathcal{L}^{(3)}$ and $\mathcal{L}^{(4)}$ with the same strength, $g^2$.

Including the effect of such interactions will lead to source terms in the RHS of eq.~\eqref{eq:Tfree},
\be
\frac{d^2T_{\pm}(\vec{k},\tau )}{k^2d\tau ^2} + \bigg[1+ \frac{2m_Q\xi}{(k\tau)^2}\mp \frac{2(\xi +m_Q)}{(-k\tau)}\bigg]T_{\pm}(\vec{k},\tau )= \mathcal{S}^{(3)}(\vec{k},\tau )+\mathcal{S}^{(4)}(\vec{k},\tau ).
\label{eq:Ts34}
\ee
From now on we will use the shorthand notation $T_k(\tau) \equiv T_{\pm}(\vec{k},\tau )$. These source terms can be calculated using the cubic and quartic Hamiltonians, defining
\be 
\mathcal{H}^{(i)}(\tau)\equiv \int d^3q\, T_q(\tau)\mathcal{S}^{(i)}(-\vec{q},\tau ) ,
\ee
from the cubic Hamiltonian~\cite{Dimastrogiovanni:2024lzj}, 
\bea
\mathcal{S}^{(3)}(-\vec{k},\tau ) &=& g\int \frac{d^3 q}{(2\pi)^{3/2}} E_3(\vec{k}, \vec{q}, \tau) T_q(\tau) T_{\vert \vec{k}-\vec{q}\vert}(\tau) ,\\
{\rm where} ~ ~ E_3(\vec{k}, \vec{q}, \tau) &=& \frac{\epsilon _1}{3} +\frac{\xi}{\tau}\frac{\epsilon _2}{3} + \frac{m_Q}{\tau}\epsilon _3 ,
\eea
where $\epsilon_1, \epsilon_2$, $\epsilon_3$ are the products of polarization vectors, given by
\begin{align}\label{ept}
\nn
\epsilon_1(\vec{q}_1,\vec{q}_2,\vec{q}_3) &= \frac{1}{2}\vec{\epsilon}_+(\vec{q}_1)\cdot\left(\vec{\epsilon}_+(\vec{q}_2)\times\vec{\epsilon}_+(\vec{q}_3)\right) \bigg\{[i(\vec{q}_3-\vec{q}_2)\cdot\vec{\epsilon}_+(\vec{q}_1)]\,[\vec{\epsilon}_+(\vec{q}_2)\cdot\vec{\epsilon}_+(\vec{q}_3)] \\ \nn
&\quad\quad\quad\quad\quad\quad\quad\quad\quad\quad\quad\quad\quad+[i(\vec{q}_2-\vec{q}_1)\cdot\vec{\epsilon}_+(\vec{q}_3)]\, [\vec{\epsilon}_+(\vec{q}_1)\cdot\vec{\epsilon}_+(\vec{q}_2)] \\ \nn
&\quad\quad\quad\quad\quad\quad\quad\quad\quad\quad\quad\quad\quad + [i(\vec{q}_1-\vec{q}_3)\cdot\vec{\epsilon}_+(\vec{q}_2)]\, [\vec{\epsilon}_+(\vec{q}_3)\cdot\vec{\epsilon}_+(\vec{q}_1)] \bigg\}, \\ \nn
     \epsilon_2(\vec{q}_1,\vec{q}_2,\vec{q}_3) &=\left[\vec{\epsilon}_+(\vec{q}_1)\cdot\left(\vec{\epsilon}_+(\vec{q}_2)\times\vec{\epsilon}_+(\vec{q}_3)\right)\right]^2, \\
    \epsilon_3(\vec{q}_1,\vec{q}_2,\vec{q}_3) &=\left[\vec{\epsilon}_+(\vec{q}_1)\cdot\vec{\epsilon}_+(\vec{q}_2)\right]\left[\vec{\epsilon}_+(\vec{q}_2)\cdot\vec{\epsilon}_+(\vec{q}_3)\right]\left[\vec{\epsilon}_+(\vec{q}_3)\cdot\vec{\epsilon}_+(\vec{q}_1)\right], 
\end{align}
and are symmetric under the exchange of any internal momenta.
Similarly, using 
\bea
\mathcal{H}^{(4)}(\tau) &=& - \mathcal{L}^{(4)}(\tau) \supset  -\frac{g^2}{4} \int   \frac{d^3 q_1 d^3 q_2 d^3 q_3 d^3 q_4}{(2\pi)^{3}} \delta ^{(3)}(\vec{q}_1 + \vec{q}_2 +  \vec{q}_3 + \vec{q}_4 ) \nonumber \\
& \times  &E_4(\vec{q}_1, \vec{q}_2 , \vec{q}_3, \vec{q}_4, \tau) T_{q_1}(\tau) T_{q_2}(\tau) T_{q_3}(\tau) T_{q_4}(\tau)\\
\mathcal{S}^{(4)}(-\vec{k},\tau ) &=& -\frac{g^2}{4} \int   \frac{d^3 q d^3 p }{(2\pi)^{3}} E_4(\vec{k}, \vec{q}, \vec{p} , \tau), \nonumber \\
&\times &T_{q}(\tau) T_{p}(\tau) T_{\vert \vec{k}-\vec{q}-\vec{p}\vert}(\tau),\\
{\rm where,}\nonumber \\
E_4(\vec{k}, \vec{q}, \vec{p} , \tau) &=& \vert \vec{\epsilon _+}(\vec{k}). \vec{\epsilon _+}(\vec{p})\vert
\vert \vec{\epsilon _+}(\vec{q}). \vec{\epsilon _+}(\vec{k}-\vec{p}-\vec{q})\vert \nonumber \\
&& \times \vert \vec{\epsilon _+}(\vec{k}). \vec{\epsilon _+}(\vec{q})\vert
\vert \vec{\epsilon _+}(\vec{p}). \vec{\epsilon _+}(\vec{k}-\vec{p}-\vec{q})\vert .\label{eq:e4}
\eea
In the next section , we will use cubic and quartic vertices to determine the non-linear sources present in the gauge field equation of motion, Eq.~\eqref{eq:Ts34}.
\section{Gauge field interactions}
\label{sec:interactions}
The gauge field interactions discussed in the previous section can lead to a large production of the $T_+$ modes. At any given time, the modes with large occupation numbers interact through the cubic and quartic self-coupling vertices through the $1+2 \leftrightarrow  3$ and $1+2 \leftrightarrow 3+4$ processes, {such that energy is carried towards the UV, from near-horizon modes to the smaller scales, deeper inside the horizon. The key contribution comes from incoming modes within the instability band, something that provides us with a natural cut-off. One has, for the zero vev case, $1\leq E_k/aH \leq 2\xi$, and for the non-vanishing vev, $x_-\leq E_k/aH  \leq x_+$.}

These interactions  allow for an exchange of energy between the modes, shifting the spectrum from the cold inflation scenario towards a possible `thermal spectrum'. 
 In contradistinction to ~\cite{Ferreira:2017lnd}, we shall use the term `efficient interaction' rather than `thermalization' to signify interactions that lead to large nonlinearities, since our definition of thermalization is more stringent.
 In the rest of this section, we attempt to derive bounds for attaining efficient interactions through two distinct approaches.
\subsection{Boltzmann equation and particle number}
\begin{figure}[hbt!]
\begin{center}
\includegraphics[width=0.5\textwidth]{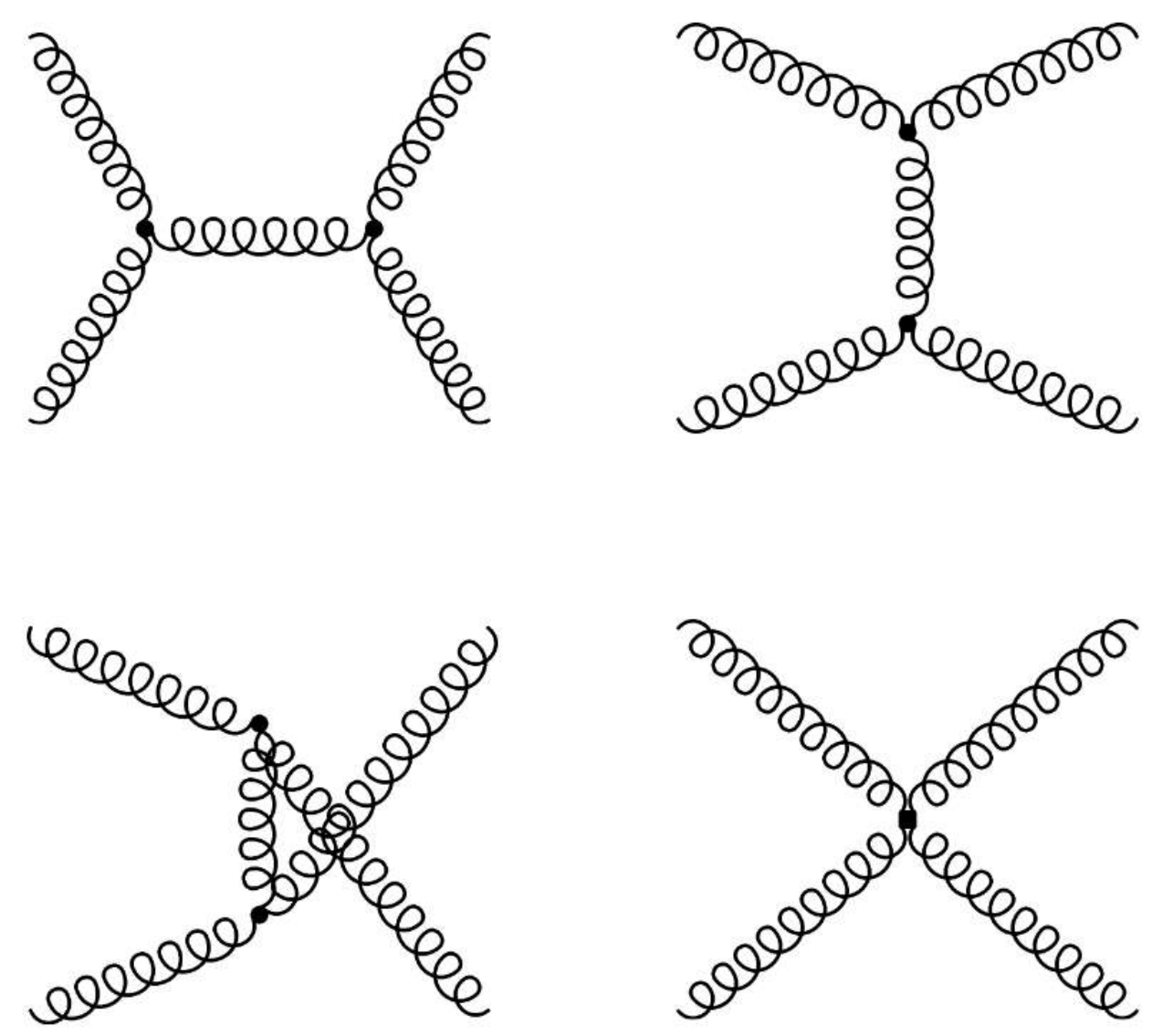}
\caption{Scattering diagrams from cubic and quartic interactions}
\label{fig:Feynmann}
\end{center}
\end{figure}
We closely follow~\cite{Ferreira:2017lnd} for the calculation of scattering diagrams. The cubic and quartic vertices under scrutiny are those in Fig.~\ref{fig:Feynmann}. For a given field $X$, with mode functions $X_k$, the comoving particle number can be defined as the ratio between energy density $\rho _k = k^2\vert X_k\vert ^2 + \vert X'_k\vert ^2$ and energy per particle $E_k$. Therefore,  
\be
N_k(\tau)+1/2 \equiv \frac{k^2\vert T_k\vert ^2 + \vert T'_k\vert ^2}{2E_k},\label{eq:comNk}
\ee
where primes denote derivatives with respect to comoving time $\tau$.
It turns out to be convenient to rewrite Eq.~\eqref{eq:Ts34} as 
 
\bea
T''_k(\tau) + \omega ^2 (\tau) T_k(\tau) &=& \mathcal{S}_k(\tau),  \\
 {\rm where }~ ~ \omega ^2 (\tau)\equiv k^2 &+& \frac{2m_Q\xi}{\tau^2} + 2\frac{(\xi + m_Q)k}{\tau},
\eea 
and $\mathcal{S}_k(\tau) \equiv \mathcal{S}^{(3)}_k(\tau) + \mathcal{S}^{(4)}_k(\tau)$ is the full source term.
Using this form and Eq.~\eqref{eq:comNk}, the corresponding Boltzmann equation is
\bea
N'_k &= & \frac{\vert T_k\vert \vert T'_k\vert }{E_k}\bigg(k^2-\omega ^2(\tau) + \frac{\mathcal{S}_k(\tau)}{\vert T_k\vert}\bigg)\nonumber \\
&=& \frac{2\vert G(k,\tau)\vert}{k^2 + \vert G(k,\tau)\vert ^2}\bigg(k^2-\omega ^2(\tau) + \frac{\mathcal{S}_k(\tau)}{\vert T_k\vert}\bigg)(N_k(\tau)+1/2)\\
{\rm where} ~ ~ G(k, \tau) &\equiv &T_k'(\tau)/T_k(\tau).\label{eq:Boltz1}
\eea

\begin{figure}[hbt!]
\begin{center}
\includegraphics[width=0.74\textwidth]{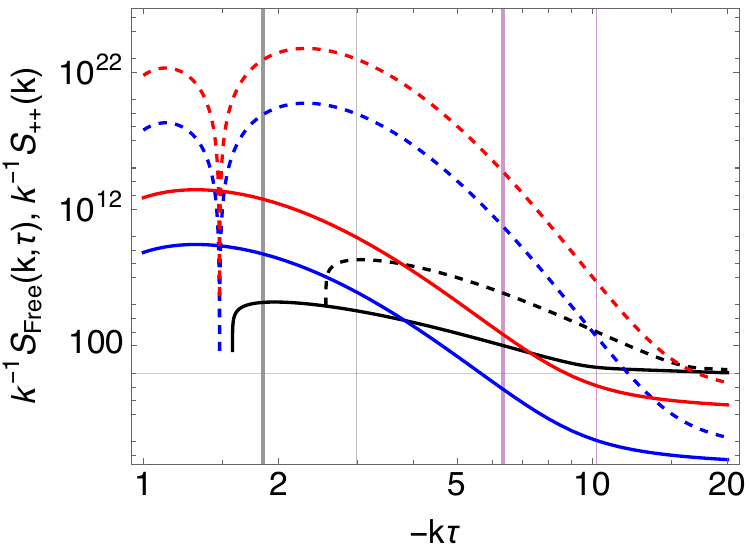}
\caption{Comparison of the source-free and sourced contributions to the Boltzmann Eq.~\eqref{eq:Boltz001}. $S_{++}(k)$ is plotted as a function of $-k\tau$ in red and blue solid (dashed) lines for $m_Q=3$ ($m_Q=5$) for $g=10^{-1}$ and $g=10^{-2}$ respectively. Black solid  (dashed) plots represent $S_{\rm Free}(k,\tau)$ for $m_Q=3$ ($m_Q=5$). Thick (thin) gray and pink vertical lines denote the position of peak $x_{-}$ and maximum instability $x_m$ for  $m_Q=3$ ($m_Q=5$).}
\label{boltzcomp}
\end{center}
\end{figure}

This first-order Boltzmann-like differential equation is exact, and it has a suitable form to interpret the scattering term $S_{++}(k)$ such that
\bea
\frac{dN_k}{d\tau}&=& \frac{2\vert G(k,\tau)\vert}{k^2 + \vert G(k,\tau)\vert ^2}(k^2-\omega ^2(\tau) )(N_k(\tau)+1/2)+ S_{++}(k)\; ,
 \label{eq:Boltz001}
\eea

\bea
S_{++}(k) &=&\frac{1}{2E_k}\int \frac{d^3p_1}{(2\pi)^{3} 2E_1} \frac{d^3p_2}{(2\pi)^{3} 2E_2} \frac{d^3p_3}{(2\pi)^{3} 2E_3} (2\pi)^4\delta ^{(4)} (k+p_1-p_2-p_3)\nonumber \\
 &&\times  \vert \mathcal{M}\vert ^2\times  \mathcal{B}(k,p_1,p_2,p_3),
 \label{eq:Boltz2}
\eea
where the scattering term is written for a $T_k + T_{p_1}\rightarrow T_{p_3} + T_{p_4}$ scattering process with energies $E_k$,$E_1$,$E_2$ and $E_3$, respectively, and $\mathcal{B}(k,p_1,p_2,p_3)$ is the phase factor, which is of  order $N_k^3$. We can immediately relate the scattering term to the source term via 
\be
S_{++}(k)= \frac{2\vert G(k,\tau)\vert}{k^2 + \vert G(k,\tau)\vert ^2}(N_k(\tau)+1/2)\frac{\mathcal{S}_k(\tau)}{\vert T_k\vert} .  \label{eq:Boltz3}
\ee
In Eq.~\eqref{eq:Boltz2} the dependence on the time variable $\tau$ is implicit in $S_{++}(k)$, when written in terms of the scattering amplitude $\vert \mathcal{M}\vert ^2$, since $E_k$ and the integration limits for the momenta depend on the time-dependent quantities $\xi$ and $x_+$. The effect of the source-free part in the right hand side of the Boltzmann Eq.~\eqref{eq:Boltz001} can be assumed to be small in the presence of efficient production of the gauge modes, since 
\be
S_{\rm Free}(k,\tau) \equiv \frac{2\vert G(k,\tau)\vert}{k^2 + \vert G(k,\tau)\vert ^2}(k^2-\omega ^2(\tau) )(N_k(\tau)+1/2)\propto N_k, 
\label{eq:Sfree}
\ee
and $S_{++}(k) \propto N_k^3$. However, the evolution of $S_{++}(k)$ and $S_{\rm Free}(k,\tau)$ are different inside the horizon and it is therefore important to specify the time $\tau$ at which their contribution is compared\footnote{Since $G(k,\tau)$ is used here only to evaluate $S_{\rm Free}(k,\tau)$, using the free solution for $T_k$ to calculate $G(k,\tau)$ is justified.}. In Fig.~\ref{boltzcomp} these are plotted for the modes at horizon, $k=aH$ (equivalently for $r\equiv aH/E_k=1$) , for which the effect of scattering is the lowest. For all the cases plotted, at the position of the peak $-k\tau = x_-$ (gray vertical lines), the scattering processes (red and blue curves) contribute more than the free term (black curves). However, if evaluated at the position of maximum instability $-k\tau = x_m=m_Q+\xi$ (pink vertical lines), for the cases such as  $m_Q=3, g=10^{-2}$, then the free term (solid black line) is comparable to the scattering term (solid red/blue line), without contributing directly to scattering processes. However, larger $m_Q$ and $g$ lead to more efficient scattering processes, and $S_{++}(k)$ contributes more than $S_{\rm Free}(k,\tau)$, even  at  $-k\tau = x_m$, which can be seen comparing the dashed red and dashed blue lines with the dashed black curve in Fig.~\ref{boltzcomp}. 

Before calculating the full effect, this comparison leads us to an understanding of the parameter space: for sufficiently large values of $g$ and $m_Q$, the scattering effects are always efficient enough to contribute to the growing particle number of smaller modes.
When the scattering effect is dominating in the Boltzmann equation, using the expression for $S_{++}(k) $ derived in Appendix~\ref{AppendixA}, efficient interaction requires that the total occupation number $N_k$ is sourced from the scattering term, i.e. $ S_{++}(k) \gg aHN_k$. This leads to the following condition for an efficient scattering process:
\be
N_k \gg  \frac{1}{2g^2\mathcal{U}^{1/2}}\bigg(\frac{aH}{E_k}\bigg)^{1/2}.
\label{eq:Nklimit1}
\ee

By efficient scattering we shall mean an interaction sourced by modes in the instability band such that also Eq. (\ref{eq:Nklimit1}) is satisfied.

For a more precise comparison, we will also include the free term in our evaluation of the above inequality. Thus, comparing $aHN_k$ with $S_{++}(k)$ and $ S_{\rm Free}(k,\tau)$, leads to a modification of the limit in Eq.~\eqref{eq:Nklimit1} as
\be
N_k \gg  \frac{1}{2g^2\mathcal{U}^{1/2}}\bigg(\frac{aH}{E_k}+\frac{S_{\rm Free}(k,\tau)}{E_k}\bigg)^{1/2}.
\label{eq:Nklimit2}
\ee
Including $S_{\rm Free}(k,\tau)$ in the comparison to determine efficiency of scattering may seem counter-intuitive, since this term does not contain interactions between different modes. However, it is still important to include this term, since it can have comparable contribution for the subhorizon modes, particularly for small to moderate values of $g$ and $m_Q$ (see Fig.~\ref{boltzcomp}). They are expected to have an important impact in the numerical evolution of the energy spectrum for a range of modes, particularly for modes in the range $x_m\leq -k\tau \leq x_+$ (or $\xi\leq -k\tau \leq 2\xi$ in the zero vev case).
\subsection{Perturbativity bounds}
\label{sec:loop}
 The strength of the gauge field self-interactions is reflected in the 3-point and 4-point vertices in Eq.~\eqref{eq:L3L4}. Efficient interactions lead to large  1-loop contributions to the gauge field power spectrum and large values for the ratio $\mathcal{R}_T $, defined as 
\be
\mathcal{R}_T \equiv  \bigg\vert \frac{P^{(1)}_{+}(k,\tau)}{P^{(0)}_+(k,\tau)}\bigg\vert \;,
\label{eq:pertratioSI}
\ee
where the power spectra are defined as 
\bea
\langle T_k(\tau) T_{k'}(\tau)\rangle_{\rm tree}&\equiv &\delta ^{(3)}(\vec{k}+\vec{k}') P^{(0)}_+(k,\tau)\nonumber \\ {\rm} ~ ~  \langle T_k(\tau) T_{k'}(\tau)\rangle_{\rm 1-loop}&\equiv &\delta ^{(3)}(\vec{k}+\vec{k}') P^{(1)}_+(k,\tau).\label{eq:defpow}
\eea
The perturbativity treatment is valid for $\mathcal{R}_T \ll 1$. As non-linear interactions become more and more important, the ratio approaches $\mathcal{R}_T \simeq 1$. 
\begin{figure}[hbt!]
\begin{center}
\includegraphics[width=0.71\textwidth]{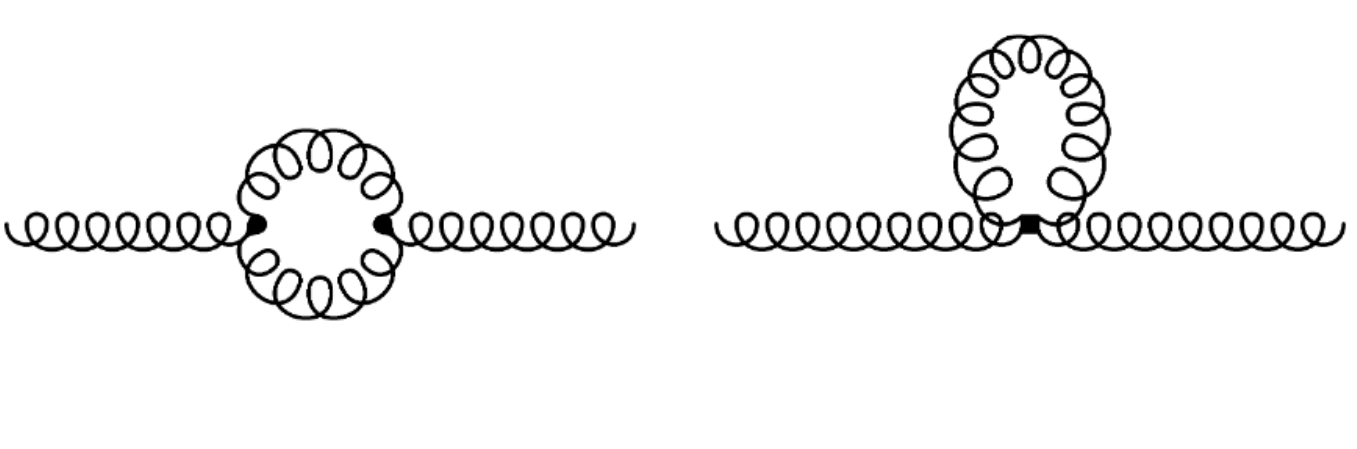}
\caption{1-loop diagrams from cubic (left) and quartic (right) interactions}
\label{fig:34ptloop}
\end{center}
\end{figure}

For the specific case of chromo-natural inflation considered here, the universality of perturbativity bounds has been discussed in~\cite{Dimastrogiovanni:2024lzj} for loops originating from $\mathcal{L}^{(3)}$ in Eq.~\eqref{eq:L3L4}, as shown in the left panel of Fig.~\ref{fig:34ptloop}. For these vertices, the 1-loop power spectrum can be evaluated using in-in formalism~\cite{Weinberg:2005vy}:
\be
\langle \hat{T}_k(\tau) \hat{T}_{k'}(\tau)\rangle^{(3)}_{\rm 1-loop} = -\int _{-\infty}^{\tau} d\tau '\int _{-\infty}^{\tau '} d\tau '' \langle [[\hat{T}_k(\tau)\hat{T}_{k'}(\tau),\hat{H}^{(3)}(\tau ')],\hat{H}^{(3)}(\tau '')]\rangle , \label{eq:inin3}
\ee
where $\hat{H}^{(3)}(\tau) =-\int d^3x \mathcal{L}^{(3)}(\tau)$. The perturbative bounds derived in this manner can be seen as identifying a specific region in the parameter space. In the chromo-natural attractor of Eq.~\eqref{eq:CNI-attractor} this corresponds to the bound~\cite{Dimastrogiovanni:2024lzj},
\be
\mathcal{R}_T =\frac{g^2}{2\pi^2}f_{2,ttt}(x_p,m_Q)\ll 1,
\label{eq:univMFlimit}
\ee
where $x_p\simeq x_-$ is the peak position of the gauge mode amplitude.
In this work we extend the above result to the case of a generic gauge field vev, and indicate  the explicit $\xi$ and $m_Q$ dependence  in the $f_2$ function as
\be
f_{2,{\rm cubic}} (x,\xi,m_Q)= \sum _{j=1,2} f^{(j)}_{2,{\rm cubic}} (x,\xi,m_Q).\label{eq:univlimgen}
\ee
The limit $m_Q=0$ leads to a pure zero vev case, whereas for the non-zero vev case, the bound in~\eqref{eq:univMFlimit} can be recovered from Eq.~\eqref{eq:univlimgen} using the chromonatural attractor limit~\eqref{eq:CNI-attractor}.
The various contributions $f^{(j)}_{2,{\rm cubic}} (x,\xi,m_Q)$ come from two distinct parts of the 1-loop power spectrum: a part that is symmetric under the  $\tau ' \leftrightarrow \tau''$ exchange, and a part that is not. They are expressed as, respectively:
\bea
\frac{f^{(1)}_{2,{\rm cubic}} (x,\xi,m_Q)}{e^{3\pi(\xi+m_Q)}} &=& \frac{1}{2} \int _0^{\infty} d\tilde{q}\int _{-1}^{1}d\eta \frac{\tilde{q}}{\tilde{p}}\bigg \vert \int _x^{x_{\rm cut}} dx' \frac{{\rm Im}[W^{\star}_{\alpha , \beta}(-2ix)W_{\alpha , \beta}(-2ix')]}{W_{\alpha , \beta}(-2ix)}\nonumber \\
&&\times \bigg[\tilde{\epsilon _1}-\frac{\xi}{x'}\epsilon _2 -\frac{3m_Q}{x'}\epsilon _3\bigg]W_{\alpha , \beta}(-2i\tilde{q}x)W_{\alpha , \beta}(-2i\tilde{p}x')\bigg\vert ^2 \nonumber \\
\frac{f^{(2)}_{2,{\rm cubic}} (x,\xi,m_Q)}{e^{3\pi(\xi+m_Q)}} &=& \int _0^{\infty} d\tilde{q}\int _{-1}^{1}d\eta \frac{\tilde{q}}{\tilde{p}}\int _x^{x_{\rm cut}} dx' \int _x^{x_{\rm cut}} dx''  \frac{{\rm Im}[W_{\alpha , \beta}(-2ix)W^{\star}_{\alpha , \beta}(-2ix')]}{\vert W_{\alpha , \beta}(-2ix)\vert ^2}\nonumber \\
&&\times \bigg[\tilde{\epsilon _1}-\frac{\xi}{x'}\epsilon _2 -\frac{3m_Q}{x'}\epsilon _3\bigg] \bigg[\tilde{\epsilon _1}-\frac{\xi}{x'}\epsilon _2 -\frac{3m_Q}{x'}\epsilon _3\bigg]^{\star}{\rm Re}[W_{\alpha , \beta}(-2ix)W^{\star}_{\alpha , \beta}(-2ix'')]\nonumber \\
&&\times {\rm Im}[W_{\alpha , \beta}(-2i\tilde{q}x')W^{\star}_{\alpha , \beta}(-2i\tilde{q}x'')W_{\alpha , \beta}(-2i\tilde{p}x')W^{\star}_{\alpha , \beta}(-2i\tilde{p}x'')]\nonumber 
\eea
where $\tilde{\epsilon}_1 \equiv \epsilon _1/k$, $\tilde{q}\equiv q/k$, $\tilde{p}\equiv p/k=\sqrt{1-2\tilde{q}\eta + \tilde{q}^2}$. The products of the polarizations $\epsilon _{1,2,3}$ are defined in Eq.~\eqref{ept} and can be expressed in terms of $\eta ,\tilde{q}$ and $\tilde{p}$, which are simply cosines of angles between the momenta. We follow standard notations here, so that the 1-loop power spectrum is computed for the external legs (Fig.~\ref{fig:34ptloop}) with momenta $k$ and $k'$, whereas the momenta  $q,p$ run inside the loop.

Similarly, the contribution from the quartic vertex (right panel of Fig.~\ref{fig:34ptloop}) of the same order in the parameter $g$ as correction to the tree-level power spectrum is
\be
\langle \hat{T}_k(\tau) \hat{T}_{k'}(\tau)\rangle^{(4)}_{\rm 1-loop} = -\int _{-\infty}^{\tau} d\tau ' 
\langle [\hat{T}_k(\tau)\hat{T}_{k'}(\tau),\hat{H}^{(4)}(\tau ')]\rangle , \label{eq:inin4}
\ee
where $\hat{H}^{(4)}(\tau) =-\int d^3x \mathcal{L}^{(4)}(\tau)$. The relevant commutators are 
\bea
\langle [\hat{T}_k(\tau)\hat{T}_{k'}(\tau), \hat{T}_1(\tau ')\hat{T}_2(\tau ')\hat{T}_3(\tau ')\hat{T}_4(\tau ')]\rangle &=& \langle [\hat{T}_k(\tau),\hat{T}_1(\tau ')]\hat{T}_k'(\tau)\hat{T}_2(\tau ') \hat{T}_3(\tau ') \hat{T}_4(\tau ')  + 3 {\rm ~ perms.} \rangle \nonumber \\
&=& \delta _{k1}\delta _{k'2}\delta_{34}(T_k(\tau)T_1^{\star}(\tau ') -{\rm c.c.})\nonumber \\
&&\times (T_{k'}(\tau)T_2^{\star}(\tau ')T_3(\tau ')T_4^{\star}(\tau ') -{\rm c.c.}) + 11 {\rm ~ perms.} \;,\nonumber \\
\eea
where  we have used the shorthand $T_i(\tau ') \equiv T_{q_i}(\tau ')$, and $\delta_{pq}$ is $\delta^{(3)}(\vec{p}-\vec{q})$. The number of permutations can be understood as  follows. The mode $k$ can contract with available momenta in 4 ways, and for each such choice, $k'$ can contract with the rest of the momenta in 3 ways, which leaves only one possibility for the internal modes to contract among themselves.
All of the permutations in this case will have identical contributions, leading to 
\bea
\langle T_k(\tau) T_{k'}(\tau)\rangle ^{(4)}_{\rm 1-loop} &=& 6g^2\int \frac{d^3q}{(2\pi)^3}\int d\tau ' \delta ^{(3)}(\vec{k}-\vec{q})\tilde{E}_4(\vec{k},\vec{q})\nonumber \\
&&\times {\rm Im}[T_k(\tau)T_k^{\star}(\tau ')]\bigg({\rm Re}[T_k(\tau)T_k^{\star}(\tau ')]{\rm Im}[T_q(\tau ')T_q^{\star}(\tau ')]\nonumber \\
&&+{\rm Im}[T_k(\tau)T_k^{\star}(\tau ')]{\rm Re}[T_q(\tau ')T_q^{\star}(\tau ')]\bigg).
\eea
Here $\tilde{E}_4(\vec{k},\vec{q}) = E_4(\vec{k},\vec{q},-\vec{q})$ can be evaluated using Eq.~\eqref{eq:e4}. One can evaluate $f_{2, {\rm quartic}}$  in a  manner similar to that used for the cubic vertices. However, this contribution has 2 fewer powers of the mode function with respect to the contribution from cubic vertices and it will therefore provide a subdominant contribution to the power spectrum. This is expected since the terms contributing to $g^2$ corrections in the power spectrum have 6 internal modes involved in each of the cubic contributions in contrast to the 4 internal modes in the quartic contribution. For this reason, we employ only the contribution from the cubic vertices in determining the perturbative bound
\be
\mathcal{R}_T 
\simeq \frac{g^2}{2\pi^2}f_{2,{\rm cubic}}(x_p,\xi,  m_Q)\ll 1.
\label{eq:loop34}
\ee

We can now evaluate bounds stemming from  perturbativity using Eq.~\eqref{eq:univlimgen} in both the zero vev and non vanishing vev (the chromo attractor) configurations. The latter was already found in~\cite{Dimastrogiovanni:2024lzj} and expressed as 
\be
\mathcal{R}^{Q\neq 0}_T=\frac{g^2}{2\pi ^2}e^{-0.074+4.25m_Q-0.00372m_Q^2} \ll 1. 
\label{eq:mqpertnAbel}
\ee
In a similar manner, for $Q=0$, we can write the perturbative bound in terms of the model parameters $\xi$ and $g$ as:
\be
\mathcal{R}^{Q=0}_T=\frac{g^2}{2\pi ^2}e^{-11.51+6.2\xi-0.035\xi^2} \ll 1 \; , 
\label{eq:pertSInew}
\ee
where the 1-loop contributions are evaluated at the horizon $x=-k\tau =1$. 

We conjecture here that violation of these perturbative limits can be used as a proxy for efficient interactions, i.e. for the regime where non-linear interactions start becoming important and can no longer be ignored. We shall see that these analytical inequalities arising from 1-loop corrections, and the bounds from scattering discussed in the previous subsection identify similar parts of the model parameter space.

\section{The warm inflation regime}
\label{sec:conditions}

\subsubsection*{Three main conditions}

 The CNI model enters a fully thermal regime when the processes leading to a sustained thermal spectrum are satisfied exponentially quickly. In particular, a `warm axion $SU(2)$ inflation' can be realized when all of the following conditions are satisfied:\\
$(i)$ Gauge fields interact efficiently, where the efficiency of nonlinear interactions is achieved by satisfying the bound  in Eq.~\eqref{eq:Nklimit1}\footnote{We show in Sec.~\ref{sec:results} that demanding $\mathcal{R}_T\simeq 1$ using Eqs.~\eqref{eq:mqpertnAbel},~\eqref{eq:pertSInew} as an alternative condition for efficient interactions identifies nearly the same parameter space.}.\\
$(ii)$ Gauge field interactions allow for a gradual modification of the energy spectrum towards a thermal spectrum for the perturbations.\\ 
$(iii)$ The thermal spectrum is sustained during inflation.

Satisfying the above conditions ensures that the gauge fields have a sustained thermal spectrum. Depending explicitly upon the strength of the Chern-Simons coupling, expected to be weaker than the gauge field self-coupling\footnote{This can be understood by comparing the number of gauge modes present in a scattering process involving a gauge field mediator  to that mediated by a scalar field (i.e. via Chern-Simons coupling). The same can also be seen in Fig.~\ref{fig:paramspace2} by comparing the perturbative bounds arising from self-interaction and Chern-Simons interaction.} for moderate to large values of $g$, 
the axion-inflaton may or may not thermalize  at a later time. Once the first condition is satisfied, the energy spectrum is modified for the modes taking part in the interaction that lay within the instability band. However, a complete transition from the cold inflation spectrum to a thermal one requires tracking the time evolution of individual modes. This involves not only accounting for the primary effect of each mode, but also considering potential secondary and tertiary interactions. For instance, a mode deep inside the horizon may contribute iteratively to shifting the overall energy spectrum before it exits the horizon. Its spectrum can be modified during its first participation in an efficient interaction, which influences subsequent evolution.

Another important matter is to rigorously account for the Hubble expansion during such process given that the expansion may hinder scattering efficiency. 
The study of the complete evolution of the spectrum until and after thermalization requires numerical analysis\footnote{Recently ref.~\cite{Kamali:2024qme} studied the effect after full thermalization, assuming that one is possible.}, ideally performed via lattice simulations. We postpone this to future work.

Whether or not the third condition is satisfied can however be estimated analytically. A thermal spectrum can sustain when the rate of thermalization is greater than the expansion rate, i.e.,  $\Delta > H$. For $SU(N_c)$ axion inflation, the rate of thermalization has been derived\footnote{We caution that reader that this and similar results found in the literature are derived under the assumption of an equal number of left and right polarized bosons. We shall use this result in the context where one polarization of gauge bosons is produced much more copiously than another. This will necessarily introduce a level of uncertainty in our results and makes them more qualitative. The natural next step is to consider numerical simulation of our setup so as not to rely on the above assumptions. We leave this to future work.} in terms of the coupling $\alpha \equiv g^2/4\pi$ and temperature~\cite{McLerran:1990de,Fu:2021jhl,Berges:2020fwq,DeRocco:2021rzv}
\be
\label{eq:rateofth}
\Delta \approx 10 N_c^2\alpha ^2 T,
\ee
with $N_c$ the number of colors in the sector. It follows that the third condition enforces a lower limit on the radiation energy density $\rho _R = g_* T^4 \pi ^2/30$:
\be
\rho _R > \frac{\pi ^2}{30}g_*\bigg(\frac{H_{\rm inf}}{10 N_c^2\alpha ^2}\bigg)^4\; ,
\label{eq:thirdconditionGraham}
\ee
where $g_*=2N_c^2 -1$.
A successful inflationary phase also requires that the total energy density $\rho _{\rm tot} \equiv \rho _R + \rho _{\rm inf} = 3H_{\rm inf}^2M_{\rm Pl}^2$ receive the main contribution from the inflaton, so that $\gamma\equiv\rho _R/\rho _{\rm tot} \ll 1 $. The third condition leads then also to a lower limit on the coupling $g$,
\be
\frac{\pi ^2}{30}g_*\bigg(\frac{H_{\rm inf}}{10 N_c^2(\frac{g^2}{4\pi})^2}\bigg)^4 \leq \gamma \times 3H_{\rm inf}^2M_{\rm Pl}^2 \; .
\label{eq:ourthirdcond}
\ee

 We shall now check the conditions for efficient interaction, Eq.~\eqref{eq:Nklimit2}, for chromo-natural inflation in two configurations: (i) vanishing gauge field vev, when the dynamics can be described in terms of the parameters $g$ and $\xi$;  (ii) non-zero gauge field vev $Q$, with the system described fully by $g$ and $m_Q$ in the regime of Eq.~\eqref{eq:CNI-attractor}. We  consider $\xi$ and $m_Q$ to be constants, as during slow-roll their values change very slowly.

 The conditions for warm inflation are evaluated at two points during the evolution of the modes: at the point of maximum instability, $-k\tau = x_m$ and the peak $-k\tau = x_p$. For the zero vev and non-zero vev cases, these points can be expressed in terms of the parameters: $x_m = \xi$, $x_p = 1$ and $x_m = \xi + m_Q$, $x_p = x_-$ respectively. The bounds are plotted with a band spanning  $r_{\rm min}\leq r \leq 1$, where $r_{\rm min}=1/2\xi\; ,$ $ x_{+}$   for the zero vev  and non-zero vev cases respectively\footnote{Although the peak for $T_k(\tau)$  occurs at $-k\tau = x_-$, including the limit upto the horizon is recommended, since the modes decrease gradually.}.

 The particle description of interactions is valid deep inside the horizon and becomes less accurate as a modes grows closer to the horizon. It is therefore more reliable at $-k\tau = x_m$  than at $-k\tau =x_p$. On the other hand, 
 bounds derived at $x_p$ make the warm regime appear slightly more likely as interactions are more efficient there. 
 We will discuss the findings corresponding to using both $x_m$ and $x_p$
 as the benchmark point. As we shall see, using the breakdown of perturbativity as a marker for the parameter space corresponding to the warm regime does not run into the same small ambiguity. 

\subsubsection*{Dynamical emergence of the vacuum expectation}
The conditions that signal the emergence of a warm regime will clearly depend on the values of the gauge field vev and are expressed as a function of the theory parameters and key time-dependent quantities. It is then crucial to keep in mind that, as a result of dynamical evolution, a non-zero vev may develop. 
Indeed, in the chromo-natural inflation scenario it is possible to dynamically evolve from a zero vev configuration towards a non vanishing one.  It was shown in ~\cite{Domcke:2018rvv} that
 two attractor solutions exist for the background: the so-called $c_0$ solution corresponding to zero vev, the other, $c_2$, emerging only for sufficiently large values of $\xi$. The work in  \cite{Domcke:2018rvv} provides clear evidence that $c_0$-type solutions become rare for large $\xi$, a regime that strongly favors $c_2$-type solutions, hence one can justifiably talk about an emerging or dynamically generated non-zero vev.
The condition identified  for the emergence of non-zero vev  is
\be
g\langle T_{\cal AB}^2\rangle ^{1/2} \simeq \frac{\xi}{-\tau},
\label{eq:emergencelimit}
\ee
where $T_{\cal AB} (k) $, defined in Eq.~\eqref{eq:Abelianlimit_Wh},
are the solutions of the gauge field fluctuations in the ``Abelian limit'' (i.e. for vanishing vev). The variance can be calculated as 
\bea
\langle T_{\cal AB}^2\rangle ^{1/2}& \equiv & \langle 0 \vert T_{\cal AB}(\tau, \vec{x}) T_{\cal AB}(\tau, \vec{x}) \vert 0 \rangle ^{1/2}\nonumber\\
&=& \bigg(\int \frac{d^3k}{(2\pi ) ^3} T_{\cal AB} (k) T_{\cal AB}^{\star} (k)\bigg)^{1/2}\nonumber \\
&=& \frac{1}{2\pi(-\tau )}e^{\pi \xi /2}\sqrt{\int _0 ^{x_{UV}} dx ~ x ~ \vert W_{-i\xi, 1/2}(-2ix) \vert ^2}.
\eea
Using $x_{UV}=2\xi$, the variance can be calculated as a function of $\tau$ for given $\xi$ so that the above condition  can be formulated in terms of $\xi$ and $g$. However, as stressed in~\cite{Domcke:2018rvv} the process of emergence is likely to take place over a time interval one cannot estimate exactly within their treatment. We can account for such an uncertainty by placing a $c_F$ factor
in front of the RHS of Eq.~\eqref{eq:emergencelimit} and let it vary within a one order of magnitude interval. One should also account for the probability of bubble nucleation from the metastable to stable vacuum. However, as shown in Appendix~\ref{AppendixD}, this rate is typically very small for the parameters under consideration.  

\subsubsection*{Steering clear of the strong backreaction regime}

Another constraint we shall implement on the parameter $\xi$ arises from requiring that we stay clear from the parameter space corresponding to strong backreaction~\cite{Peloso:2016gqs} of the gauge fields on the scalar field dynamics. 
Using the definition of $\xi$, this can be written as
\be
\frac{e^{\pi \xi}}{\xi ^{5/2}} \ll 158 \frac{1}{\lambda}\frac{f}{H}\; .
\label{eq:xibackAbel}
\ee
For vanishing gauge field vev, this is the only constrain stemming from avoiding strong backreaction. In the $Q\neq 0$ case, a more stringent limit originates from avoiding the regime of strong backreaction of the gauge field perturbations on the $Q$ background equation of motion.
A semi-analytical bound in terms of the parameters $g$ and $m_Q$~\cite{Dimastrogiovanni:2016fuu,Papageorgiou:2019ecb,Dimastrogiovanni:2024lzj} takes the following form
\be
g\ll \bigg(\frac{24\pi ^2}{2.3e^{3.9m_Q}}\frac{m_Q}{m_Q+1/m_Q}\bigg)^{1/2}.
\label{eq:mqbacknAbel}
\ee

\subsubsection*{Perturbativity}
We find it worthwhile here to briefly review the bounds on the parameter space of axion - gauge field models stemming from perturbativity. The reason they are of interest in this work is the fact  that these identify a map of a parameter space qualitatively very similar to the one that will emerge from  efficient scattering. The bounds on perturbativity~\cite{Ferreira:2015omg,Peloso:2016gqs,Dimastrogiovanni:2024lzj} stem from requiring that the one loop contribution to the gauge field two-point function is smaller than its tree-level counterpart.  The ratio under scrutiny is defined as 
\be
\mathcal{R}^{(\alpha)}_T \equiv  \bigg\vert \frac{P^{(1)}_{+,\alpha}(k,\tau)}{P^{(0)}_+(k,\tau)}\bigg\vert ,
\ee

where the power spectra are those in Eq.~\eqref{eq:defpow}. 
The superscript/subscript $\alpha$ indicates the specific vertex responsible for the loop. The resulting inequality can be   written in terms of the particle production parameters.\\ For the zero vev case, the interactions in the loop originate from the Chern-Simons term $\sim \phi\,T_{\cal AB}^2$ and the perturbative bound is~\cite{Peloso:2016gqs}
\be
\mathcal{R}^{\phi\,T_{\cal AB}^2}_T \simeq e^{2.01\pi(\xi-4.6)} \ll 1.
\label{eq:xipertAbel}
\ee
 Due to the (finite) tachyonic instability in the gauge sector, we expect the perturbativity bounds obtained  from gauge fields self-interaction to be more stringent on the parameter space. Such bounds on $\mathcal{R}_T \ll 1$ can be readily expressed in terms of the model parameters discussed in section~\ref{sec:loop}. In addition to that in Eq.\eqref{eq:xipertAbel},  we will employ the  bounds given in Eqs.~\eqref{eq:pertSInew} and~\eqref{eq:mqpertnAbel} for zero and non-vanishing vev respectively.

 \bigskip
  To summarize, besides the three conditions for a warm inflation regime, we will have to engage with  (a) the fact that the inflationary potential ought to give the leading contribution to the energy density; (b) the dynamical emergence of a non-zero vev for large values of $\xi$; (c) the constraints stemming from avoiding the strong backreaction regime. In this section, we also reviewed perturbativity constraints to facilitate later comparison with requirements of a different origin.

\section{Results and Discussion}
\label{sec:results}

We shall now illustrate the various regions in the $(g,\xi)$ plane corresponding to efficient scattering and sustained thermalization. In the same plane we superimpose the areas corresponding to strong backreaction, perturbativity, and the emergence of the vev. We then do the same for the case of non-zero vev.

\begin{figure}[hbt!]
\includegraphics[ width=0.49\textwidth]{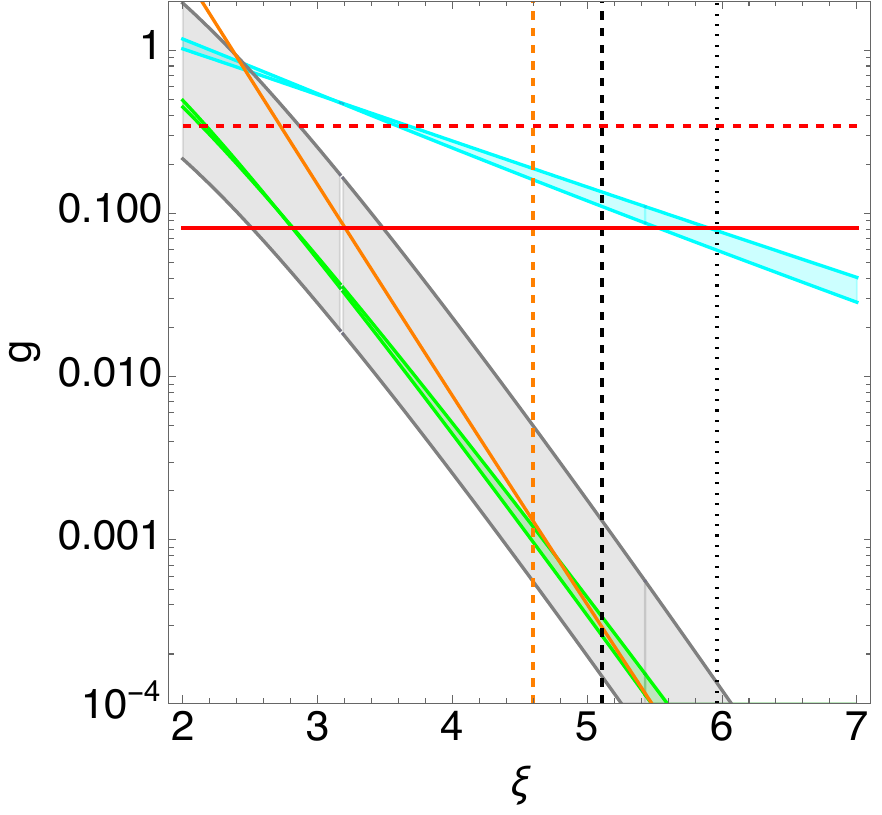}
\includegraphics[ width=0.495\textwidth]{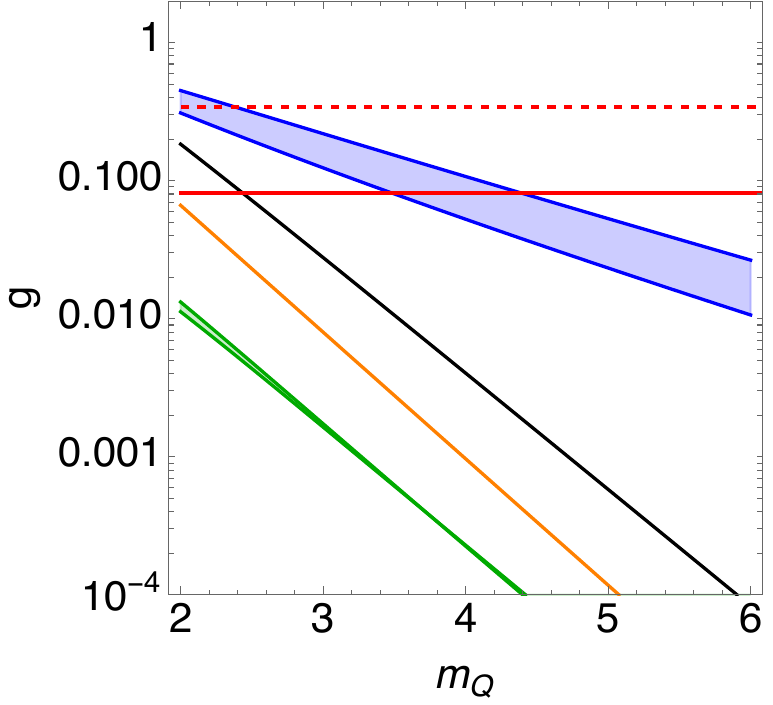}
\caption{The various regions in the $(\xi,g)$ plane identifying the conditions for transitioning to the warm inflation regime. \textit{Left:} Bounds for zero vev case.  Interactions are efficient above the light green (cyan) band plotted for $1/2\xi <r<1$, where the gauge modes are estimated at the horizon exit $x=1$ (at maximum instability $x=\xi$). Backreaction bounds are shown in black dashed  and dotted vertical lines for $\frac{1}{\lambda}\frac{f}{H}=10^3 $ and $10^4$ respectively. Perturbativity bounds from the Chern-Simons contribution and from self-interaction are shown in orange dashed and solid lines respectively. \textit{Right:} Bounds for non-vanishing vev.  Interactions are efficient above the dark green (blue) band plotted for $x_+ <r<1$, where the gauge modes are estimated at the peak $x=x_-$ (at maximum instability $x=x_m$). In both  panels, red horizontal solid (dashed) lines signify sustainability of thermalization with $H_{\rm inf}=10^{-10}~M_{\rm Pl}$ ($H_{\rm inf}=10^{-5}~ M_{\rm Pl}$).}
\label{fig:paramspace2}
\end{figure}

The left and right panel of Fig.~\ref{fig:paramspace2} refer to the vanishing and non-zero vev case respectively. In both panels $N_k(\tau)$ is evaluated at the horizon  (left: light green band; right: dark green band) and at the position of maximum instability  (cyan and blue bands). 
 Condition (i) toward warm inflation (see Section \ref{sec:conditions}) is satisfied above these bands, where the interaction among the gauge modes is efficient. 
 Once above these bands, we assume for simplicity  that condition (ii) for warm inflation is satisfied immediately.
 Ours is merely a simplifying assumption, a more in-depth numerical study is called for to fully justify this choice. We leave it to future work. Condition (iii) takes the form  of horizontal solid and dashed red lines indicating the minimum $g$ value necessary for a sustained thermalization (i.e. Eq.~\eqref{eq:ourthirdcond}) with $\gamma = 0.5$ for, respectively, $H_{\rm inf}=10^{-10}M_{\rm Pl}$ and $H_{\rm inf}=10^{-5}M_{\rm Pl}$.  

The warm regime of axion - $SU(2)$ inflation is then confined to a parameter space characterized by $\mathcal{O}(0.1 - 1)$ values of the coupling (the precise value depending on $H_{\rm inf}$),  a result we arrived at more heuristically when comparing $S_{\rm Free}$ and $S_{++}$ in Fig.~\ref{boltzcomp}. The parameter space below the light green  band in the left panel and below dark green  band in the right panel can thus be safely considered a region of  cold axion inflation  (with zero and non-zero vev respectively).

In the left panel, we show with a gray band the parameter space satisfying  Eq.~\eqref{eq:emergencelimit}, the condition for a transition to a non-zero vev. The $c_{F}$ factor, limited to the interval $1/3\leq c_F \leq 3$,  accounts for the inaccuracy in determining the exact transition line. 
 The black dashed and dotted vertical lines in this panel represent the strong backreaction bounds on $\xi$ for $\frac{1}{\lambda}\frac{f}{H} =10^3\; ,10^4$ respectively (see Eq.~\eqref{eq:xibackAbel}). Operating in the weak backreaction regime excludes the region to the right of these vertical lines. The dashed orange vertical line, obtained via Eq.~\eqref{eq:xipertAbel},  signals that perturbativity is violated for large $\xi$ values.
 The other perturbativity bound, stemming from gauge field self-interactions (c.f. Eq.~\eqref{eq:pertSInew}), is represented by a solid orange line.

 For the non-vanishing vev case (right panel), the backreaction bound corresponding to Eq.~\eqref{eq:mqbacknAbel} is represented by a solid black line while the inequality in Eq.~\eqref{eq:mqpertnAbel} is to be identified with the orange line.

Considering the left panel of Fig.~\ref{fig:paramspace2}
 when the interaction strength is evaluated at horizon crossing, we see that efficient interactions are possible above the green band, and sustained thermalization is in place for $g$ above the red line (solid or dashed one, depending on the value of $H_{\rm inf}$). Warm axion inflation in the case of vanishing  vev is then possible in the parameter space above the efficient interaction bound (green band) and the sustained thermalization line.
 
 One should also keep in mind the parameter space identified by the criterion for the emergence of a non zero vev, represented by a gray band. One should move to the right panel of Fig.~\ref{fig:paramspace2} once the evolution of the initially-zero-vev system reaches the gray region. The gray band corresponding to the emergence of a vev fully overlaps, at least in region of parameter space in hand, with the green line signaling strong non-linearities, thus emphasizing the necessity for a thorough numerical lattice-based analysis of the system.
 The diagonal line identifying the perturbativity bound from gauge field self-interactions is most interesting. It is remarkable how closely it lies to the green band associated with efficient interactions. This suggests that perturbativity is, in this case, a good proxy for condition (i) stemming from the Boltzmann treatment of scattering processes. Even in the case of non-zero vev (right panel of Fig.~\ref{fig:paramspace2}) the perturbativity bound is off but only a factor of a few with respect to the line identified by condition (i) and it is, to a good approximation,  parallel to it in the $(q,m_Q)$ plane.
 
 Another notion to consider at this stage is the fact that the onset of efficient interactions, the vev emerging parameter space, and the strong backreaction bound, have all been arrived at perturbatively, in particular employing the cold-regime wavefunction.  The exact results one may pursue via lattice simulations may then deviate from our findings above the orange perturbativity bound. Nevertheless, even in such region ours remains  a reliable qualitative  picture one can use as a useful guide to chart the parameter space where warm inflation can occur. 
 
If, in the pursuit of efficient interactions, instead of evaluating $N_k(\tau)$ at the horizon we do so at the point of maximum instability, the region where condition (i) is satisfied is typically significantly above the perturbativity bound.

\section{Conclusions and relation to previous work}
\label{sec:conclusions}
Axion inflation is a compelling class of models for early-time acceleration. The (approximately) shift-symmetric potential tackles the $\eta$-problem and can be embedded in UV complete approaches such as string theory \cite{Baumann:2014nda}. Coupling the axion sector to gauge field is rather natural in that it does not break the shift symmetry and it is often implemented by the dimension 5 operator known as  Chern-Simons term. The ensuing signatures can be very intriguing, from chiral gravitational waves to primordial black holes, all these attributes being testable by next-generation cosmological probes.

Given that many of the more interesting observational features of this class of models can be traced back to the non-linearities in the gauge sector, it is important to explore the possibility that a thermal regime may emerge and lead to modified predictions.
 We address this question for a particular setup, where the axion inflaton is coupled to an $SU(2)$ gauge sector. We start out in a cold inflation regime and investigate the possibility of dynamically transitioning into the warm paradigm.  

We identified three necessary conditions for the transition and studied the associated parameter space. We have addressed condition (ii) only argumentatively and assumed it to be verified essentially instantaneously with condition (i). We postpone a more in-depth analysis for a future work, which will include the full numerical (or stochastic) calculation required.

We found the first condition is satisfied for large $\mathcal{O}(0.1-1)$ values of gauge field self-coupling $g$, and for moderate to large value of $\xi$ (with $Q=0$) and $m_Q$ (for $Q\neq 0$). We also identified the bounds on $g$ stemming from the third  condition, showing that it is more easily satisfied (i.e. lower values of $g$ are acceptable) for low values of the Hubble rate during inflation (i.e. lower scale inflation). With all the bounds in hand, we found that for the zero vev case the $g-\xi$ parameter space where warm inflation may occur is rather limited. Similar conclusions apply to the $Q\neq 0$ case.

We have superimposed the necessary conditions for a warm regime with that of requiring perturbative control over the model, finding a remarkable overlap between perturbativity bounds and those stemming from condition (i). As a caveat, one should remark that the perturbativity bounds were derived under the assumption of slowly rolling $\xi, m_Q$, which allows analytical estimates. A numerical analysis would provide a more accurate derivation of the bound and its hierarchy with respect to the strong backreaction bound. Indeed, in our analysis we demanded that the system evolution stays away from the strong backreaction regime. So long as perturbativity is not violated, the strong backreaction regime can be well-described by existing analytical and  numerical techniques. We leave this step for future work. In our parameter space plots, we have also accounted for the prospect \cite{Domcke:2018rvv} of a dynamically emerging vev. We calculated the probability associated with the alternative  path to a non-zero vev, quantum tunneling finding it to be negligibly small.

\subsubsection*{Relation to previous work}
Regarding previous work on the topic, we are indebted to the
authors of~\cite{Ferreira:2017lnd},  whose work explored the particle scattering approach for axion inflation models with Abelian gauge fields.  Our description of the non-Abelian gauge field sub-horizon modes as particles closely follows their approach, and we explicitly explore self-interaction, which turns out to be the dominant contribution for CNI models. 
The minimal warm inflation model with $SU(3)$ gauge fields presented in ~\cite{Berghaus:2019whh} has a small initial temperature, i.e., weak thermalization is assumed as an initial condition, whereas in this work we have studied the emergence of thermalization with a cold initial condition. Reference~\cite{Mukuno:2024yoa} also worked under the assumption of an initially thermalized configuration and explored the values of the sphaleron transition coefficient~\cite{Laine:2021ego} for entering a strong warm CNI regime. The work in~\cite{Fujita:2025zoa} considered reheating in axion inflation and provided a schematic flowchart of different conditions necessary for the strong warm inflation regime. In the context of this flowchart, we have investigated the parameter space for leading to warm inflation in the absence of an initial thermal bath. 

In~\cite{DeRocco:2021rzv} a  CNI-like model with $SU(3)$ gauge fields is explored and thermalization conditions are derived.
In view of the first condition presented in~\cite{DeRocco:2021rzv}, we have investigated here whether the non-linearities become relevant within a Hubble time (our condition 1) for both the vanishing and non-zero vev case. In our work we also stressed that, although large non-linearities mark a definite departure from the cold spectrum, the drift to a thermal spectrum within a Hubble time is not guaranteed and calls for further analysis. 
The studies in~\cite{DeRocco:2021rzv,Berghaus:2019whh,Mukuno:2024yoa} work with $\mathcal{O}(0.1)$ values of $\alpha$, corresponding to large values of the self-coupling $g$, which indeed belong to the parameter space we found to have the optimal conditions for thermalization.
Moreover, we find that the cold regime of CNI is in place for a very large fraction of the parameter space. Our work here bridges the gap in the literature by considering a dynamically evolving and a non-zero vacuum expectation value in the non-Abelian case. 

\acknowledgments 
We thank Carlos Pena, Juan Garcia-Bellido, and Martino Michelotti for illuminating discussions. We thank Simona Procacci for insightful discussions. SB,  MF, and AP acknowledge the ``Consolidaci\'{o}n Investigadora'' grant CNS2022-135590. The work of SB, MF, and
AP is partially supported by the Spanish Research Agency (Agencia Estatal de Investigaci\'{o}n)
through the Grant IFT Centro de Excelencia Severo Ochoa No CEX2020-001007-S, funded by
MCIN/AEI/10.13039/501100011033. MF acknowledges support from the ``Ram\'{o}n y Cajal'' grant
RYC2021-033786-I. Part of this work was carried out during the 2025 ``The Dawn of Gravitational Wave Cosmology'' workshop, supported by the Fundaci\'{o}n Ram\'{o}n Areces and hosted by the ``Centro de Ciencias de Benasque Pedro Pascual''. We thank both the CCBPP and the Fundaci\'{o}n Areces for creating a stimulating and very productive environment for research.

\newpage
\appendix
\section{Scattering contribution}
\label{AppendixA}
To calculate the contribution of the scattering term $S_{++}(k)$ in terms of the parameters and amplitude of scattering, we note that $\delta ^{(4)} (k+p_1-p_2-p_3) = \delta ^{(3)} (k+p_1-p_2-p_3)\delta (E_k+E_1-E_2-E_3)$, and $ d^3p_i = \vert p_i \vert ^2 dp_id\Omega = \vert p_i \vert  E_idE_id\Omega $. In the center of mass (CoM) frame, $\vec{k} = -\vec{p}_1$, so that the 3D delta function above leads to $\vec{p}_3 = -\vec{p}_2$. Also, the Mandelstam variable $s\equiv (k+p_1)_{\mu}(k+p_1)^{\mu} = (p_2+p_3)_{\mu}(p_2+p_3)^{\mu} = (E_k + E_1)^2 = (E_2 + E_3)^2$. In this frame, for massless particles, 
$E_k = E_1 = E_2 = E_3 = \frac{\sqrt{s}}{2}$, so that $\delta (E_k+E_1-E_2-E_3) =  \frac{1}{2}\delta (E_3 -  \frac{\sqrt{s}}{2})$.\\
 We can relate $S_{++}$ in the Boltzmann equation Eq.~\eqref{eq:Boltz2} to the scattering cross-section. 
\bea
S_{++}(k) &=& \int \frac{d^3p_1}{(2\pi)^{3} } \langle \sigma v_{\rm rel} \rangle , ~ ~ {\rm where }\\
\langle \sigma v_{\rm rel} \rangle &=& \frac{1}{4E_kE_1}\int \frac{d^3p_2}{(2\pi)^{3} 2E_2}\int \frac{d^3p_3}{(2\pi)^{3} 2E_3} (2\pi)^4\delta ^{(4)} (k+p_1-p_2-p_3) \nonumber \\
&&\times \vert \mathcal{M}\vert ^2\times  \mathcal{B}(k,p_1,p_2,p_3).
\eea
Using CoM frame, thus, 

\begin{equation}
    \langle \sigma v_{\rm rel} \rangle = \frac{ 1}{64\pi ^2 E_k E_1 \sqrt{s}}\int \vert\vec{p}_2\vert\, dE_2 \;\delta  \left(E_2-\frac{\sqrt{s}}{2}\right) \mathcal{B}(k,p_i) \int d\Omega  \vert \mathcal{M}\vert ^2,
\end{equation}
and,  $\vert\vec{p}_2\vert = E_2$ leads to
\be
\langle \sigma v_{\rm rel} \rangle =  \frac{\mathcal{B}(k,p_i)}{64\pi ^2 E_k E_1}\frac{1}{2} \int d\Omega  \vert \mathcal{M}\vert ^2,
\ee
The phase space factor in this $1+2\rightarrow 3+4$ scattering process is $\mathcal{B}(k,p_i)\equiv N_kN_{p_1}(1+N_{p_2})(1+N_{p_3})-N_{p_2}N_{p_3}(1+N_k)(1+N_{p_1})$. The integral peaks for a range of modes close to $k$, with limits defined by the model parameters $\xi$ or $m_Q$. Therefore, the leading contribution from the phase factor can be estimated as $\mathcal{B}(k,p_i)\simeq N_k^3\equiv \mathcal{B}(k)$. So, with $\vert\vec{p}_1\vert = E_1$,
\bea
S_{++}(k) &=&  \frac{ \mathcal{B}(k)}{2.(2\pi)^3.64\pi ^2 E_k }\int dE_1 E_1 \int d\Omega \int d\Omega \vert \mathcal{M}\vert ^2 \nonumber \\
&=& \mathcal{B}(k) \frac{s}{E_k}g^4\mathcal{U} = 4\mathcal{B}(k) E_k g^4 \mathcal{U}, ~ ~ {\rm where}\\
\mathcal{U}&\equiv& \frac{1}{16\times 64\pi ^3}\int d(\cos \theta) \vert \mathcal{M}\vert ^2.
\eea 
 In this setup, we will work in the CoM frame with the following descriptions of the momenta and polarization vectors.
 \bea
 k^{\mu} = (E,0,0,E),  ~ ~ ~ ~ ~ ~ &&p_1^{\mu} = (E,0,0,-E),\nonumber \\
p_2^{\mu} = (E,E\sin \theta,0,E\cos \theta), ~ ~ ~ ~ ~ ~ 
&&p_3^{\mu} = (E,-E\sin \theta,0,-E\cos \theta),\nonumber \\
\epsilon_{\pm}^{\mu} (k) = \frac{1}{\sqrt{2}}(0,1,\pm i,0), ~ ~ ~ ~ ~ ~ 
&& \epsilon_{\pm}^{\mu} (p_1) = \frac{1}{\sqrt{2}}(0,-1,\pm i,0),\nonumber \\
\epsilon_{\pm}^{\mu \star} (p_2) = \frac{1}{\sqrt{2}}(0,-\cos \theta,\pm i,\sin \theta), ~ ~ ~ ~ ~ ~ 
&& \epsilon_{\pm}^{\mu \star}(p_3) = \frac{1}{\sqrt{2}}(0,\cos \theta,\pm i,-\sin \theta).
\eea
\section{Scattering amplitudes}
\label{AppendixC}
The vertices coming from the cubic and quartic Lagrangian that involve the largest number of gauge fields are:
\bea
\mathcal{L}^{(3)}& \supset & -gf^{abc}A_{ai}A_{bj}\partial _iA_{cj} - \frac{g\xi }{3\tau} f^{abc}\epsilon ^{ijk}A_{ai}A_{bj}A_{ck} -\frac{gm_Q}{\tau}A_{ij}A_{jk}A_{ki}  \\
\mathcal{L}^{(4)} &\supset & -\frac{1}{4}g^2f^{abc}f^{ade}A_{bi}A_{cj}A_{di}A_{ej} .
\eea
at this point, we choose to exploit the fact that the structure constants for the $SU(2)$ algebra are $f^{abc} = \epsilon ^{abc}$. With this, and reminding ourselves that $A_{ii}=0$, the chromo term can be rewritten as 
\be
-\frac{gm_Q}{\tau}A_{ij}A_{jk}A_{ki} = \frac{gm_Q}{\tau}\delta_{ci}\epsilon ^{abe}\epsilon ^{cde} A_{ai}A_{bj}A_{dj}.
\ee 
\subsubsection*{Contribution 1 }
From $-gf^{abc}A_{ai}A_{bj}\partial _iA_{cj} $ in $\mathcal{L}^{(3)}$, 
\bea
-gA_{ai}(\vec{A}_j \times \partial _i \vec{A}_j)_a &=& -g[A_{ix}(A_{jy}\partial _iA_{jz} - A_{jz}\partial _iA_{jy}) +  A_{iy}(A_{jz}\partial _iA_{jx} - A_{jx}\partial _iA_{jz}) \nonumber \\
&+&  A_{iz}(A_{jx}\partial _iA_{jy} - A_{jy}\partial _iA_{jx}) ]
\eea
so that the contribution from a vertex with legs $A_{la}$,$A_{mb}$ and $A_{kc}$ with incoming momenta $p_1$,$p_2$ and $p_3$ respectively,  will be 
\be
-ig[\delta_{lm}(\vec{p}_1-\vec{p}_2)_k+ \delta_{mk}(\vec{p}_2-\vec{p}_3)_l + \delta_{kl}(\vec{p}_3-\vec{p}_1)_m].
\ee
The contributions from the s-, t- and u- channel are
\bea
i\mathcal{M}_s &=& g^2\cos \theta,\\
i\mathcal{M}_t &=& -g^2\frac{1}{4(1-\cos \theta)}\bigg[17+3\cos \theta -13 \cos ^2\theta + \cos ^3\theta \bigg],\\
i\mathcal{M}_u &=& -g^2\frac{1}{4(1+\cos \theta)}\bigg[17-3\cos \theta -13 \cos ^2\theta - \cos ^3\theta \bigg]
\eea
so that 
\bea
\mathcal{M}_{I} &=& ig^2\bigg[A_s \cos \theta - A_t \frac{1}{4(1-\cos \theta)}\bigg(17+3\cos \theta -13 \cos ^2\theta + \cos ^3\theta \bigg) \nonumber \\
&-& A_u \frac{1}{4(1+\cos \theta)}\bigg(17-3\cos \theta -13 \cos ^2\theta - \cos ^3\theta \bigg)\bigg].
\eea
Evaluating this way, $A_s = A_t = A_u = 1$.

\subsubsection*{Contribution 2}
From $\frac{g\xi }{3\tau} f^{abc}\epsilon ^{ijk}A_{ai}A_{bj}A_{ck} $ in $\mathcal{L}^{(3)}$, with $g'=\frac{1}{3}g\xi aH = \frac{g\xi}{-3\tau}$, 
\bea
-\frac{g\xi }{3\tau} f^{abc}\epsilon ^{ijk}A_{ai}A_{bj}A_{ck} &=& -g'\epsilon^{ijk}[A_{ix}(A_{jy}A_{kz} - A_{jz}A_{ky}) +  A_{iy}(A_{jz}A_{kx} - A_{jx}A_{kz}) \nonumber \\
&+&  A_{iz}(A_{jx}A_{ky} - A_{jy}A_{kx}) ]
\eea
so that the contribution from a vertex with legs $A_{la}$,$A_{mb}$ and $A_{nc}$ with incoming momenta $p_1$,$p_2$ and $p_3$ respectively,  will be 
\be
-g'\epsilon^{ijk}[\delta_{il}(\delta _{jm}\delta _{kn} - \delta _{jn}\delta _{km} ) +  \delta_{im}(\delta _{jn}\delta _{kl} - \delta _{jl}\delta _{kn} ) + \delta_{in}(\delta _{jl}\delta _{km} - \delta _{jm}\delta _{kl} ) ].
\ee
The contributions from the s-, t- and u- channel are
\bea
i\mathcal{M}_s &=& g^2\xi ^2\bigg(\frac{aH}{E}\bigg)^2 \cos \theta,\\
i\mathcal{M}_t &=& \frac{1}{2}g^2\xi ^2\bigg(\frac{aH}{E}\bigg)^2 \bigg(\frac{3+2\cos \theta -\cos ^2\theta}{1-\cos \theta}\bigg),\\
i\mathcal{M}_u &=& \frac{1}{2}g^2\xi ^2\bigg(\frac{aH}{E}\bigg)^2 \bigg(\frac{3-2\cos \theta -\cos ^2\theta}{1+\cos \theta}\bigg)
\eea
so that 
\be
\mathcal{M}_{II} = -ig^2\xi ^2\bigg(\frac{aH}{E}\bigg)^2 \bigg[B_s \cos \theta + B_t\frac{3+2\cos \theta -\cos ^2\theta}{2(1-\cos \theta)} +B_u\frac{3-2\cos \theta -\cos ^2\theta}{2(1+\cos \theta )} \bigg].
\ee
Evaluating this way, $B_s = B_t = B_u = 1$.
\subsubsection*{Contribution 3}
$-\frac{gm_Q }{\tau} A_{ij}A_{jk}A_{ki} $ in $\mathcal{L}^{(3)}$ can also be written as
\bea
\frac{gm_Q }{\tau}\epsilon ^{abe}\epsilon ^{cde} \delta _{ci} A^a_i A^b_j A^d_j &=& -\frac{gm_Q }{\tau}[A_{ix}A_{jx}(A_{jy} + A_{jz}) + A_{iy}A_{jy}(A_{jz}  +  A_{jx}) \nonumber \\
&+& A_{iz}A_{jz}(A_{jx} + A_{jy})  ]
\eea
so that the contribution from a vertex with legs $A_{la}$,$A_{mb}$ and $A_{nc}$ with incoming momenta $p_1$,$p_2$ and $p_3$ respectively,  will be 
\be
-\frac{gm_Q }{\tau}[\delta_{il}\delta _{jl}(\delta _{jm} + \delta _{jn}) +  \delta_{im}\delta _{jm}(\delta _{jn} + \delta _{jl}) + \delta_{in}\delta _{jn}(\delta _{jm} + \delta _{jl}) ].
\ee
The contributions from the s-, t- and u- channel are
\bea
i\mathcal{M}_s &=& 9g^2m_Q ^2\bigg(\frac{aH}{E}\bigg)^2 \cos \theta,\\
i\mathcal{M}_t &=& \frac{9}{2}g^2m_Q ^2\bigg(\frac{aH}{E}\bigg)^2 \bigg(\frac{3+2\cos \theta -\cos ^2\theta}{1-\cos \theta}\bigg),\\
i\mathcal{M}_u &=&  \frac{9}{2}g^2m_Q ^2\bigg(\frac{aH}{E}\bigg)^2 \bigg(\frac{3-2\cos \theta -\cos ^2\theta}{1+\cos \theta}\bigg)
\eea
so that 
\be
\mathcal{M}_{III} = -9ig^2m_Q ^2\bigg(\frac{aH}{E}\bigg)^2 \bigg[C_s \cos \theta + C_t\frac{3+2\cos \theta -\cos ^2\theta}{2(1-\cos \theta)} +C_u\frac{3-2\cos \theta -\cos ^2\theta}{2(1+\cos \theta )} \bigg].
\ee
Evaluating this way, $C_s = C_t = C_u = 1$.
\subsubsection*{Contribution 4 }
$-\frac{1}{4}g^2f^{abc}f^{ade}A_{bi}A_{cj}A_{di}A_{ej} $ in $\mathcal{L}^{(4)}$ can also be written as
\bea
\frac{g^2 }{4}\vert \vec{A}_i\times \vec{A}_j \vert ^2&=& -\frac{g^2 }{4} [\vert A_{iy}A_{jz} - A_{iz}A_{jy}\vert ^2 +\vert A_{iz}A_{jx} - A_{ix}A_{jz}\vert ^2 \nonumber \\
&+& \vert A_{ix}A_{jy} - A_{iy}A_{jx}\vert ^2 ]
\eea
so that the contribution from a vertex with legs $A_{lb}$,$A_{mc}$, $A_{dn}$ and $A_{qe}$ with incoming momenta $p_1$,$p_2$, $p_3$ and $p_4$ respectively,  will be 
\bea
&&-g^2[(\delta_{il}\delta _{jm}-\delta_{im}\delta _{jl})(\delta_{in}\delta _{jq}-\delta_{iq}\delta _{jn}) \nonumber \\
&+& (\delta_{il}\delta _{jn}-\delta_{in}\delta _{jl})(\delta_{iq}\delta _{jm}-\delta_{im}\delta _{jq})  \nonumber \\
&+& (\delta_{iq}\delta _{jl}-\delta_{il}\delta _{jq})(\delta_{im}\delta _{jn}-\delta_{in}\delta _{jm}) ].
\eea
The contribution is
\be
\mathcal{M}_{IV} = -ig^2D_4(3+2\cos \theta - \cos ^2 \theta).
\ee
Evaluating this way, $D_4 = 1$.
The full scattering amplitude is therefore,
\bea
\mathcal{M} \equiv \sum _{j} \mathcal{M}_j &=& -ig^2\bigg[-cos \theta + \frac{1}{4(1-\cos \theta)}\bigg(17+3\cos \theta -13 \cos ^2\theta + \cos ^3\theta \bigg) \nonumber \\
&+& \frac{1}{4(1+\cos \theta)}\bigg(17-3\cos \theta -13 \cos ^2\theta - \cos ^3\theta \bigg) + (3+2\cos \theta - \cos ^2 \theta)\nonumber \\
&+& (h^2+9l^2)\bigg( \cos \theta + \frac{3+2\cos \theta -\cos ^2\theta}{2(1-\cos \theta)} \nonumber \\
&+& \frac{3-2\cos \theta -\cos ^2\theta}{2(1+\cos \theta )} \bigg) \bigg]
\eea
where $h^2 \equiv \xi ^2(\frac{aH}{E})^2$ and $l^2 \equiv m_Q ^2(\frac{aH}{E})^2$.

\subsection*{Limits of integration}
The propagator has at least a momentum $q/a = H$. From the $t-$ channel  and $u$-channel respectively,
\bea
\vec{q}_{t} &=& \vec{p}_1 - \vec{p}_3 = -E\sin \theta \hat{x} + E(1-\cos \theta )\hat{z}\\
\vec{q}_{u} &=& \vec{p}_1 - \vec{p}_4 = E\sin \theta \hat{x} + E(1+\cos \theta )\hat{z}.
\eea 
Demanding $\vert \vec{q}_{t,u}\vert \geq aH$, it can be shown that,
\be
\frac{r^2}{2}-1 \leq \cos \theta \leq 1- \frac{r^2}{2},
\ee
 where $r$ is defined in the main text.

\section{Efficiency of bubble nucleation for emergence of non-zero gauge field background}
\label{AppendixD}

We dedicate this Appendix to studying the possibility of the non-zero gauge field background forming from a first order phase transition. during inflation. This is in contrast to the conventional way as in \cite{Domcke:2018rvv,Domcke:2019lxq} where the non-zero background arises kinematically through the effects of perturbations on the background. From the equation of motion for the background eq.~(\ref{eq:eqQ}) we can straightforwardly write an expression for the effective potential of the gauge field in the exact deSitter approximation
\begin{equation}
    V_{\rm eff}(Q)= H^2 Q^2-\frac{2}{3}\,g\, H\, \xi\, Q^3+\frac{g^2}{2}\,Q^4
\end{equation}
This effective potential features two minima as long as $\xi>2$.
\begin{equation}
    Q_{\rm min,0}=0\;\;\;{\rm and}\;\;\; Q_{\rm min,2}=\frac{H \xi}{2 g}+\frac{H \sqrt{\xi^2-4}}{2 g}\,.
\end{equation}
The first minimum always corresponds to an effective potential of zero whereas the second one is equal to 

\begin{equation}
    V(Q_{\rm min,2})=-\frac{H^4}{12 g^2}\left(6-6\,\xi^2+\xi^4-4\,\xi\,\sqrt{\xi^2-4}+\xi^3\,\sqrt{\xi^2-4}\right)
\end{equation}
The expression above can be recast into a condition for $Q_{\rm min,0}$ being the false vacuum and $Q_{\rm min,2}$ being the true vacuum. The condition is simply $\xi>\frac{3}{\sqrt{2}}$. This implies that as long as the axion rolls down the potential fast enough there will be a possibility of the gauge field background tunneling from the false to the true vacuum.

The bubble nucleation efficiency for a first order phase transition in a quartic scalar potential has been computed in a generic manner in reference \cite{Adams:1993zs}. We define the bubble nucleation efficiency as
\begin{equation}
    \epsilon_B=\frac{\Delta V}{H^4} {\rm e}^{-S_{E4}}\;.
    \label{eq:bub-nuc}
\end{equation}
Where $\Delta V$ is the overall change in the energy scale between the two vacua, $H$ is the scale factor and $S_{E4}$ is the Euclidean action in four dimensions. Reference \cite{Adams:1993zs} has computed the result for a generic potential of the type
\begin{equation}
    V(\phi)=\lambda \phi^4-a \phi^3+b \phi^2+c\phi+d
\end{equation}
where all the coefficients of $\phi$s are understood to be constants. Mapping this language into our case, we find that 
\begin{equation}
 \lambda=  \frac{g^2}{2}\;\;\;,\;\;\;a=\frac{2}{3}\,g\,H\,\xi\;\;\;,\;\;\;b=H^2\;\;\;,\;\;\;c=d=0\;.
\end{equation}
We also define the auxiliary quantity $\delta=8\lambda b/a^2$. The final result for the Euclidean action has been computed numerically in \cite{Adams:1993zs} and fitted to
\begin{equation}
    S_{\rm E4}=\frac{\pi^2}{3 \lambda}(2-\delta)^{-3}\left[\alpha_1\delta+\alpha_2\delta^2+\alpha_3\delta^3\right]
\end{equation}
where the constants are $\alpha_1=13.832$, $\alpha_2=-10.819$ and $\alpha_3=2.0765$.

\begin{figure}[hbt!]
\centering
\includegraphics[ width=0.7\textwidth]{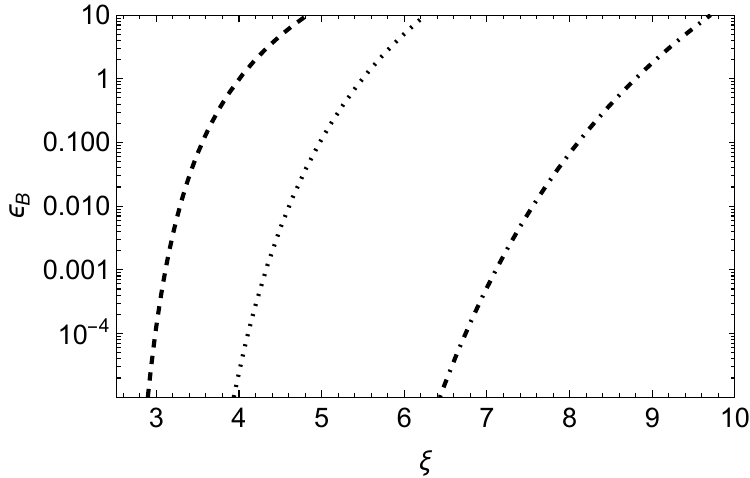}
\caption{Plot of the bubble nucleation efficiency defined in (\ref{eq:bub-nuc}) as a function of the particle production parameter $\xi$ for three different choices of the gauge coupling $g$. The dashed line corresponds to $g=10^0$, dotted line to $g=10^{-1/3}$ and dot-dashed to $g=10^{-2/3}$.}
\label{fig:bubble-nuc}
\end{figure}

Equipped with the above, we can compute the bubble nucleation efficiency in terms of two as of yet undetermined constants $g$ and $\xi$. We display in figure \ref{fig:bubble-nuc} the bubble nucleation efficiency for three distinct values of $g$ in terms of $\xi$. Comparing this plot to the main results in figure \ref{fig:paramspace2} we see that the kinematic way is always much more efficient than the first order phase transition studied here. Therefore we focus in the main text in the case where the non-zero gauge field background is accessed kinematically instead of quantum mechanically.

\newpage

\bibliographystyle{utphys}
\bibliography{refwarm}
\end{document}